\documentclass[twocolumn,showpacs,prb,aps,superscriptaddress]{revtex4-1}
\usepackage[latin1]{inputenc}
\usepackage{bm}
\usepackage{multirow}
\usepackage{amssymb}
\usepackage{amsbsy}
\usepackage{amsmath}
\usepackage{stmaryrd}
\usepackage{graphicx}
\usepackage{epsfig}

\newcommand{\op}[1]{#1}
\newcommand{\Jec}{J_\mathrm{ec}}

\begin{document}

\title{Decay of currents for strong interactions}

\author{Robin Steinigeweg}%
\email{r.steinigeweg@tu-bs.de}%
\affiliation{Institute for Theoretical Physics, Technical University Braunschweig, D-38106 Braunschweig, Germany}%
\date{\today}

\begin{abstract}
The decay of current autocorrelation functions is investigated for
quantum systems featuring strong `interactions'. Here, the term
interaction refers to that part of the Hamiltonian causing the
(major) decay of the current. On the time scale before the (first)
zero-crossing of the current, its relaxation is shown to be well
described by a suitable perturbation theory in the lowest orders of
the interaction strength, even and especially if interactions are
strong. In this description the relaxation is found to be rather
close to a Gaussian decay and the resulting diffusion coefficient
approximately scales with the inverse interaction strength. These
findings are also confirmed by numerical results from exact
diagonalization for several one-dimensional transport models
including spin transport in the Heisenberg chain w.r.t.~different
spin quantum numbers, anisotropy, next-to-nearest-neighbor
interaction, or alternating magnetic field; energy transport in the
Ising chain with tilted magnetic field; and transport of excitations
in a randomly coupled modular quantum system. The impact of these
results for weak interactions is finally discussed.
\end{abstract}

% 05. Statistical physics, thermodynamics, and nonlinear dynamical systems
% 05.30.-d Quantum statistical mechanics
% 05.60.Gg Quantum transport
% 05.70.Ln Nonequilibrium and irreversible thermodynamics


\maketitle

% ----------------------------------------------------------------------------
%
% Section: Introduction
%
% ----------------------------------------------------------------------------

\section{Introduction}
\label{introduction}

Perturbation theory is one of the main approaches to many-particle
physics with a wide range of applications in the context of quantum
transport, ranging from the investigation of Green's functions using
Feynman graphs, the setup of a (quasi-)particle description by means
of a Boltzmann equation, the derivation of the Green-Kubo formula
within linear response theory \cite{kubo1991,mahan2000}, to many
other applications in this context \cite{sirker2009}. A particular
application is the use of different projection operator techniques
\cite{nakajima1958,zwanzig1960,mori1965,forstner1975,chaturvedi1979,breuer2007}
for the realization of steady-state bath scenarios
\cite{mejiamonasterio2007,michel2008,prosen2009,steinigeweg2009-1,prosen2010,prosen2011,znidaric2011}
or the analysis of current autocorrelations
\cite{jung2006,jung2007,steinigeweg2010-2}. A perturbation theory is
commonly employed due to the availability of a small parameter,
e.g., weak particle-particle interactions, external scattering
centers, or system-bath coupling. But additional assumptions are
often required, such as the random phase and Markov approximation
\cite{mahan2000,breuer2007}.
\\
The concrete choice of a projection operator technique is a subtle
task, whenever the addressed dynamics becomes non-Markovian and
features memory effects \cite{steinigeweg2007}, typically occurring
at short time scales. Because the relevant time scales are short in
the case of strong perturbations, such non-Markovian effects appear
in an already difficult case for any perturbation theory. However,
the decay of the spin-current in the anisotropic Heisenberg chain
\cite{zotos2003,heidrichmeisner2007} at high temperatures has been
well described for the case of large anisotropy parameters in
Ref.~\onlinecite{steinigeweg2010-2} by a lowest order prediction in
the anisotropy, as obtained from a certain variant of projection
operator techniques. Moreover, the resulting quantitative values for
the diffusion coefficient have been brought into good agreement with
numerical findings in the literature
\cite{prelovsek2004,michel2008,prosen2009,huber1969-1,huber1969-2,karadamoglou2004}.
But the perturbation theory in Ref.~\onlinecite{steinigeweg2010-2}
has focused on a single quantum model so far and has been carried
out numerically on the basis of finite systems solely, leaving the
origin of the observed agreements and disagreements as an open
issue. Hence, one main intention of this paper is the extension of
the perturbation theory to a wider class of quantum models and the
analytical treatment of strong perturbations in the thermodynamic
limit (and weak perturbations close to that limit). Furthermore,
criteria for the validity of the lowest order prediction will be
formulated and higher order corrections will be taken into account.
By the use of these criteria and the comparison with numerically
exact diagonalization (ED) the relaxation of the current is found to
be well described in the lowest orders of the perturbation, even and
especially if the perturbation is not weak. In particular the
relaxation is rather close to a Gaussian decay and the resulting
diffusion coefficient roughly scales with the inverse perturbation
strength.

This paper is structured as follows: In the next Sec.~\ref{current}
the general definition of the current and the connection between its
autocorrelation and the diffusion coefficient is briefly reviewed at
first. Then the perturbation theory for the decay of the current
autocorrelation is introduced in Sec.~\ref{lowest} and the validity
of the lowest order truncation for strong perturbations is discussed
in detail here. In the following Secs.~\ref{modular}-\ref{Ising} the
introduced perturbation theory is applied to several one-dimensional
transport models in the limit of high temperatures, namely, the
transport of excitations in a randomly coupled modular quantum
system (Sec.~\ref{modular}); spin transport in the Heisenberg chain
w.r.t.~anisotropy, next-to-nearest neighbor interactions, different
spin quantum numbers, or a staggered magnetic field
(Sec.~\ref{Heisenberg}); and energy transport in the Ising chain
with a tilted magnetic field (Sec.~\ref{Ising}). The last
Sec.~\ref{summary} closes with a summary and conclusion.

% ----------------------------------------------------------------------------
%
% Section: Current and Diffusion Coefficient
%
% ----------------------------------------------------------------------------

\section{Current and Diffusion Coefficient}
\label{current}

In the present paper several (quasi-)one-dimensional and
translationally invariant quantum systems will be studied, described
by a respective Hamiltonian $\op{H}$. For such systems a globally
conserved transport quantity $\op{X}$ will be considered, i.e.,
$[\op{H}, \op{X}] = 0$. The latter quantity, and the Hamiltonian as
well, are both decomposable into $N$ local portions $\op{x}_r$ and
$\op{h}_r$, corresponding to different spatial positions $r$:
\begin{equation}
\op{X} = \sum_{r=1}^N \op{x}_r \, , \; \op{H} = \sum_{r=1}^N
\op{h}_r \, . \label{H}
\end{equation}
Here, the $\op{x}_r$ may be defined either exactly on the position
of the $\op{h}_r$, in between, or both. The above decomposition is
further done in such a way that Heisenberg's equation of motion is
of the form ($\hbar \equiv 1$)
\begin{eqnarray}
\frac{\text{d}}{\text{d}t} \, \op{x}_r = \imath [\op{H},\op{x}_r]
&=& \imath [\op{h}_{r^-},\op{x}_r] + \imath [\op{h}_{r^+},\op{x}_r]
\nonumber\\
&\equiv& \op{j}_{r-1} - \op{j}_r \, , \label{continuity}
\end{eqnarray}
where $\op{h}_{r^-}$ and $\op{h}_{r^+}$ are located directly on the
l.h.s.~and r.h.s.~of $\op{x}_r$, respectively. Because only the
contributions from next neighbors $r^-$ and $r^+$ are involved, this
form may require the choice of a proper elementary cell. For
instance, if additional contributions from next-to-nearest neighbors
occur, a larger cell consisting of two or even more sites may be
chosen. However, once a description in terms of Eqs.~(\ref{H}) and
(\ref{continuity}) has been established, the local current is
\emph{consistently} defined by $\op{j}_r = \imath
[\op{x}_r,\op{h}_{r^+}]$ \cite{steinigeweg2009-3} and the total
current reads
\begin{equation}
\op{J} = \sum_{r=1}^N \op{j}_r \, .
\end{equation}

This paper will focus on the current autocorrelation function $C(t)
= \langle J(t) J(0) \rangle$. Here, the time arguments of operators
have to be understood w.r.t.~the Heisenberg picture and the angles
denote the equilibrium average at infinite temperature, i.e.,
essentially the trace operation: $\langle \ldots \rangle =
\text{Tr}\{$\ldots \}$ / \text{dim} {\cal H}$. Particularly, the
time-integral
\begin{equation}
{\cal D}(t) = \frac{1}{\langle \op{X}^2 \rangle} \int_0^t \!
\text{d}t' \, C(t') \label{D}
\end{equation}
will be of interest. Apparently, for $t \rightarrow \infty$, the
quantity ${\cal D}(t)$ coincides with the diffusion constant
according to linear response theory \cite{kubo1991,mahan2000}.
However, for any finite time, this quantity is also connected to
the actual expectation value of local densities
\begin{equation}
d_r(t) = \text{Tr}\{ \op{x}_r(t) \rho(0) \} - \langle \op{x}_r
\rangle \, ,
\end{equation}
where $\rho(0)$ represents an initial density matrix, featuring an
inhomogeneous nonequilibrium density profile at the beginning.
Concretely, the above quantity ${\cal D}(t)$ and the spatial
variance
\begin{equation}
\text{Var}_r(t) = \sum_{r=1}^N d_r(t) \, r^2 - \left[ \sum_{r=1}^N
d_r(t) \, r \right ]^2  \label{var1}
\end{equation}
are connected via the relation \cite{steinigeweg2009-3}
\begin{equation}
\frac{\text{d}}{\text{d}t} \, \text{Var}_r(t) = 2 \, {\cal D}(t) \,
. \label{var2}
\end{equation}
Thus, whenever ${\cal D}(t)$ is constant at a certain time scale,
$\text{Var}_r(t)$ increases linearly at that scale, as expected for
the case of diffusive dynamics. Contrary, ${\cal D}(t) = 0$ yields
no increase (insulating behavior) and ${\cal D}(t) \propto t$ leads
to a quadratic increase (ballistic behavior).
\\
Strictly speaking, the relation in Eq.~(\ref{var2}) is only fulfilled
\emph{exactly} for a class of initial states $\rho(0)$
\cite{steinigeweg2009-3}, representing however an ensemble average
w.r.t.~typicality \cite{goldstein2006,popescu2006,reimann2007} or,
more accurately, the dynamical typicality of quantum expectation
values \cite{bartsch2009}. Hence, the overwhelming majority of all
possible initial states $\rho(0)$ is nevertheless expected to yield
roughly a spatial variance corresponding to the ensemble average, if
only the dimension of the relevant Hilbert space is sufficiently
large. The latter largeness is certainly satisfied for most practical
purposes. In other words, a concrete initial state $\rho(0)$ is
expected to fulfill approximately the relation in Eq.~(\ref{var2}),
at least if $\rho(0)$ is not constructed explicitly to violate this
relation. Note that the relation in  Eq.~(\ref{var2}) remains still
reasonable for lower temperatures, if all trace operations are
performed including the statistical operator at a given temperature,
see Ref.~\onlinecite{steinigeweg2009-3} for details.

% ------------------------------- FIGURE1 ------------------------------------
\begin{figure}[tb]
\centering
\includegraphics[width=0.8\columnwidth]{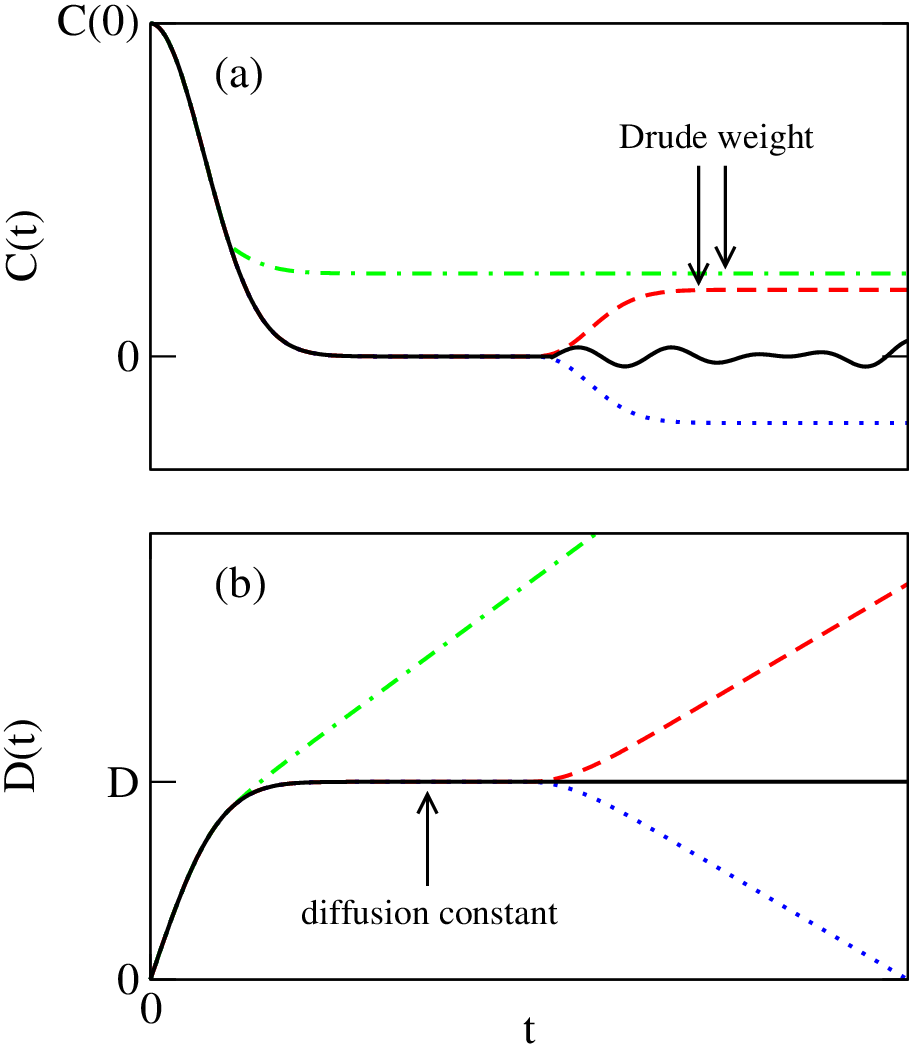}
\caption{(color online) Sketch for the possible time-dependence of
(a) the current autocorrelation function $C(t)$ and (b) its
time-integral ${\cal D}(t)$. In (a) the current autocorrelation
function $C(t)$ may decay completely at short times and then remain
zero at intermediate times (black, solid curve). At long times
$C(t)$ may decrease further to negative values (blue, dotted curve)
or instead also increase to a positive value, due to a finite Drude
weight (red, dashed curve). Such a Drude weight can already become
dominant before the zero-crossing (green, dashed-dotted curve). In
(b) the resulting time-integral ${\cal D}(t)$ may develop a plateau
at intermediate times (black, solid curve), indicating diffusive
dynamics. In that case the height of the latter plateau corresponds
to the quantitative value of the diffusion constant $\cal D$. The
width of this plateau, and its existence as such, both depend on the
concrete form of the underlying current autocorrelation function in
(a) (other curves).}  \label{sketch}
\end{figure}
%----------------------------------------------------------------------------

Generally, the dynamical behavior crucially depends on the
considered time scale. At sufficiently short times transport
\emph{generically} is ballistic: The time-integral ${\cal D}(t)$
firstly increases linearly, because the underlying current
autocorrelation function $C(t)$ has not decayed yet, at least not
significantly, see Fig.~\ref{sketch}. But, if the current is not
strictly conserved, its autocorrelation function $C(t)$ can further
decay and may eventually reach a value close to zero at intermediate
times. If $C(t)$ additionally remains at a value in the vicinity of
zero, its time-integral ${\cal D}(t)$ approximately becomes constant
and finally develops a plateau at such time scales, see
Fig.~\ref{sketch}. Then transport is diffusive and the respective
diffusion coefficient is given by the height of the latter plateau.
One main intention of this paper is to deduce the quantitative value
of this diffusion coefficient perturbatively, especially in the case
of strong perturbations, see the following Sec.~\ref{lowest}.
\\
However, even if ${\cal D}(t)$ features a plateau at intermediate
times, the dynamics does \emph{not necessarily} stay diffusive for
arbitrary long times. First, if the current is partially conserved,
its autocorrelation function $C(t)$ contains a non-decaying
contribution (finite Drude weight). Such a contribution, independent
of its weight, eventually leads to a renewed linear increase of
${\cal D}(t)$ for sufficiently long times and consequently yields
ballistic dynamics again, cf.~Fig.~\ref{sketch} (red, dashed curve).
Particularly, ${\cal D}(t)$ diverges for $t \rightarrow \infty$. The
latter similarly happens, whenever the current autocorrelation
function $C(t)$ is (quasi-)periodic in $t$ (finite recurrence time).
But the \emph{renewed} increase of ${\cal D}(t)$ in
Fig.~\ref{sketch} (red, dashed curve) typically turns out to be a
finite size effect, see Sec.~\ref{Heisenberg} and
Refs.~\onlinecite{steinigeweg2009-2, steinigeweg2010-2}.
\\
Second, apart from transitions towards ballistic behavior in the
limit of long times, transitions towards insulating behavior can
appear, of course. In that case the current autocorrelation function
$C(t)$ crosses zero and takes on a negative value, yielding a
decrease of its time-integral ${\cal D}(t)$, see Fig.~\ref{sketch}
(blue, dotted curve). However, all those transitions, occurring
after a zero-crossing of the current, are not in the focus of this
paper. In fact, the introduced perturbation theory in the following
Section addresses the dynamics on the time scale before any
occurrence of a zero-crossing. Strictly speaking, the theory itself
does not even allow for a definite conclusion on the time scale after
the first zero-crossing. But the theory yields at least an educated
guess for ${\cal D}(t)$,  if $C(t)$ is assumed to be more or less
`well-behaved', i.e., similar to Fig.~\ref{sketch} (black, solid
curve). For instance, minor unsystematic fluctuations of the current
autocorrelation function $C(t)$ around zero do not contribute
significantly to its time-integral ${\cal D}(t)$ any further.
Therefore, using this assumption, the diffusion constant is mainly
set by the time scale before the first zero-crossing. For most of
the concrete transport models in Secs.~\ref{modular}-\ref{Ising} this
assumption is indeed supported by the numerical results from ED.

% ----------------------------------------------------------------------------
%
% Section: Lowest Orders Truncations
%
% ----------------------------------------------------------------------------

\section{Lowest Orders Truncations for Strong Perturbations}
\label{lowest}

In this Section the perturbation theory for the analysis of the
dynamical behavior of the current autocorrelation function $C(t)$ is
introduced. However, this theory is not restricted exclusively to
the autocorrelation function of the current and may be applied
analogously to any other observable of interest. Moreover, even the
decomposition of the Hamiltonian according to Eq.~(\ref{H}),
considered for the definition of the current only, is not necessary
in the following.
\\
Generally, a strategy for the perturbative description of the
dynamical behavior of an autocorrelation function, such as $C(t)$,
is the application of projection operator techniques, particularly,
the time-convolutionless (TCL) method
\cite{chaturvedi1979,breuer2007,steinigeweg2010-2}, as used here. This
particular method, and the well-known Nakazima-Zwanzig (NZ) method
\cite{nakajima1958,zwanzig1960} as well, are
both applied commonly in the context of open quantum systems. In the
context of closed quantum systems the Mori-Zwanzig memory matrix
formalism \cite{mori1965,forstner1975,jung2006,jung2007} is applied
more commonly. The latter formalism is, however, similar to the NZ
method.

For the application of projection operator techniques a suitable
projection (super-)operator $\cal P$ has to be defined firstly,
projecting a density matrix $\rho(t)$ onto a relevant subspace. This
relevant subspace has to include at least the identity $\op{1}$ and
the observable of interest $\op{J}$. Thus, a natural choice of $\cal
P$ is given by
\begin{equation}
{\cal P} \, \rho(t) = \frac{1}{\text{dim}{\cal H}} + \frac{\langle
\op{J} \, \rho(t) \rangle}{\langle \op{J}^2 \rangle} \, \op{J} +
\sum_i \frac{\langle \op{A}_i \, \rho(t) \rangle}{\langle \op{A}_i^2
\rangle} \, \op{A_i} \, . \label{projector}
\end{equation}
Here, the $\op{A}_i$ denote additional observables which are of
interest by themselves or crucially affect the dynamical behavior of
the single observable of interest $\op{J}$. Without loss of
generality, the set of all operators $(\op{1}, \op{J}, \op{A_i})$
may be assumed to be orthogonal w.r.t.~the trace operation, i.e.,
$\langle J \rangle = \langle \op{A}_i \rangle = \langle \op{J} \,
\op{A}_i \rangle = \langle \op{A}_i \, \op{A}_j \rangle = 0$. If $J$
represents the current, $\op{1}$ and $\op{J}$ are always orthogonal,
since no current flows in equilibrium, i.e., $\langle \op{J} \rangle
= 0$. This orthogonality and the normalization factors in
Eq.~(\ref{projector}) guarantee ${\cal P}^2 = {\cal P}$, i.e., $\cal
P$ satisfies the property of a projection. Moreover, for initial
conditions $\rho(0)$ in the span of $\op{1}$ and $\op{J}$, the two
properties $(1 - {\cal P}) \rho(0) = 0$ and $\langle \op{J} \, {\cal
P} \, \rho(t) \rangle \propto C(t)$ are fulfilled. Especially the
latter property demonstrates the connection to the autocorrelation
function $C(t)$.

Apart from the definition of an appropriate projection
(super-)operator, the application of projection operator techniques
requires the decomposition of the Hamiltonian into the form $\op{H}
= \op{H}_0 + \Delta \, \op{V}$ with $\op{H}_0$ as the uncoupled
system and with $\op{V}$ as the `interaction'. The parameter
$\Delta$ is the strength of the interaction. This decomposition is
independent from the one in Eq.~(\ref{H}), considered for the
definition of the current solely. The decomposition of $\op{H}$ is
usually done in such a way that the interaction strength $\Delta$ is
a small parameter and the observables of interest are strictly
conserved in the uncoupled system, e.g., $[ \op{H}_0, \op{J} ] = 0$
\cite{jung2006,jung2007}. However, if the interaction strength
$\Delta$ is a \emph{large} parameter, it will be sufficient to
assume that the dynamics in $\op{H}_0$ is much slower than in
$\Delta \, \op{V}$. For \emph{very} strong interactions also the
eigensystem of $\op{H}_0$ will not be necessary. But for weak
interactions the eigensystem is indispensable and has to be found
analytically or at least numerically.

Once the projection has been chosen and the uncoupled system has
been identified, the TCL formalism routinely yields a closed and
time-local differential equation for the evolution of ${\cal P} \,
\rho_\text{I}(t)$ in the interaction picture \cite{breuer2007}
\begin{equation}
\frac{\partial}{\partial t} \, {\cal P} \, \rho_\text{I}(t) = {\cal
G}(t) \, {\cal P} \, \rho_\text{I}(t) + {\cal I}(t) \, (1-{\cal P})
\, \rho(0) \label{TCL}
\end{equation}
and avoids the often troublesome time-convolution which appears,
e.g., in the context of the NZ method. For initial conditions
$\rho(0)$ with $(1-{\cal P}) \, \rho(0) = 0$ the inhomogeneity on
the r.h.s.~of Eq.~(\ref{TCL}) vanishes, e.g., for the considered
$\rho(0)$ in the span of $\op{1}$ and $\op{J}$. The generator ${\cal
G}(t)$ is given as a systematic perturbation expansion in powers of
the interaction strength $\Delta$, namely,
\begin{equation}
{\cal G}(t) = \sum_{i = 1}^\infty \Delta^i \, {\cal G}_i(t) \, , \;
{\cal G}_{2i-1}(t) = 0 \, . \label{g}
\end{equation}
The odd contributions of this expansion vanish in many situations
and also for all concrete transport models in the subsequent
Sections. Hence, the lowest nonvanishing contribution is the second
order ${\cal G}_2(t)$. It reads, using the Liouville operator ${\cal
L}(t) = -\imath [\op{V}_\text{I}(t), \rho_\text{I}(t)]$,
\begin{equation}
{\cal G}_2(t) = \int_0^t \! \text{d}t_1 \, {\cal P} \, {\cal L}(t)
\, {\cal L}(t_1) \, {\cal P} \, . \label{g2}
\end{equation}
The next lowest nonvanishing contribution is the fourth order ${\cal
G}_4(t)$. It is given by
\begin{eqnarray}
{\cal G}_4(t) \! &=& \! \int_0^t \text{d}t_1 \int_0^{t_1}
\text{d}t_2 \int_0^{t_2} \text{d}t_3 \nonumber \\[0.1cm]
&& \! {\cal P} \, {\cal L}(t) \, {\cal L}(t_1) \, (1 - {\cal P})
\, {\cal L}(t_2) \, {\cal L}(t_3) \, {\cal P} \nonumber \\[0.1cm]
&-& \! {\cal P} \, {\cal L}(t) \, {\cal L}(t_2) \, {\cal P} \,
{\cal L}(t_1) \, {\cal L}(t_3) \, {\cal P} \nonumber \\[0.1cm]
&-& \! {\cal P} \, {\cal L}(t) \, {\cal L}(t_3) \, {\cal P} \,
{\cal L}(t_1) \, {\cal L}(t_2) \, {\cal P} \, . \label{g4}
\end{eqnarray}

Generally, Eq.~(\ref{TCL}) leads to a rate \emph{matrix} equation
for the evolution of the expectation values $\langle \op{J} \, {\cal
P} \, \rho_\text{I}(t) \rangle$ and $\langle \op{A}_i \, {\cal P} \,
\rho_\text{I}(t) \rangle$ of all operators in the chosen projection.
For the remainder, however, the discussion will focus on the
projection onto a single operator $J$. In that case Eq.~(\ref{TCL})
yields a rate equation for the dynamical behavior of the expectation
value $\langle \op{J} \, {\cal P} \, \rho_\text{I}(t) \rangle \propto
\langle \op{J}(t) \, \op{J}_\text{I}(t) \rangle$ only. As long as the
time evolution of $J$ w.r.t.~$\op{H}_0$ is negligibly slow, say,
$\op{J}_\text{I}(t) \approx \op{J}$, the latter expectation value is
identical to the autocorrelation function $C(t)$. Hence, Eq.~(\ref{TCL})
can be rewritten as
\begin{equation}
\frac{\text{d}}{\text{d}t} \, C(t) = -[\Delta^2 \, R_2(t) +
\Delta^4 \, R_4(t) + \ldots] \, C(t) \label{decay}
\end{equation}
with scalar rates $\Delta^2 R_2(t)$ and $\Delta^4 R_4(t)$,
resulting from Eqs.~(\ref{g})--(\ref{g4}). Concretely, the second
order rate $R_2(t)$ is given by
\begin{equation}
R_2(t) = \int_0^t \! \text{d}t_1 \, f(t_1) \, , \; f(t_1) =
\frac{\langle \imath[\op{J},\op{V}]_\text{I}(t_1) \,
\imath[\op{J},\op{V}] \rangle}{\langle \op{J}^2 \rangle} \label{r2}
\end{equation}
and the fourth order rate $R_4(t)$ reads
\begin{eqnarray}
R_{4}(t) &=& \int_{0}^{t} \! \text{d}t_1 \int_{0}^{t_1} \!
\text{d}t_2 \int_{0}^{t_2} \text{d}t_3 \nonumber \\[0.1cm]
&& \! f(t-t_1) \; f(t_2-t_3) \nonumber \\[0.1cm]
&+& \! f(t-t_2) \; f(t_1-t_3) \nonumber \\[0.1cm]
&+& \! f(t-t_3) \; f(t_1-t_2) \nonumber \\[0.1cm]
&-& \! \frac{\langle [ [\op{J}, \op{V}_\text{I}(t_1)],
\op{V}_\text{I}(t) ] \, [ [\op{J}, \op{V}_\text{I}(t_3)],
\op{V}_\text{I}(t_2) ] \rangle}{\langle J^2 \rangle} \, . \label{r4}
\end{eqnarray}
The above Eqs.~(\ref{decay})-(\ref{r4}) build the framework for the
analysis of the decay of the current autocorrelation $C(t)$ in the
next Sections.

So far, the rate equation is formally exact, at least if the rates
in \emph{all} orders of the perturbation expansion are taken into
account. But already the concrete evaluation of the fourth order
rate is typically a highly nontrivial task, both analytically and
numerically. The rest of this Section will therefore discuss the
possibility of a second order truncation in detail. The quality of
this truncation may crucially depend on the choice of the projection
and is commonly expected to be justified only in the limit of weak
interactions \cite{breuer2007}.
\\
In the limit of weak interactions the truncation to lowest order
usually relies on the fact that decay times become arbitrary long,
if only the strength of the interaction is sufficiently small. In
the case of long time scales, e.g., in the Markovian limit the rates
in all orders are \emph{assumed} to take on constant values
$\Delta^i \, R_i$. Due to the independence of the rates from time,
the contribution of the $i$th order rate is essentially determined
by the overall scaling factor $\Delta^i$ solely. Therefore, in the
limit $\Delta \rightarrow 0$, the dominant contribution is given by
the lowest order rate $\Delta^2 \, R_2$. By the use of this
assumption, Eq.~(\ref{decay}) directly predicts the purely
exponential decay $C(t) = \exp(-\Delta^2 R_2 \, t) \, \langle J^2
\rangle$. For times above the decay time $\tau = 1/(\Delta^2 R_2)$
the diffusion coefficient according to Eq.~(\ref{D}) consequently
becomes
\begin{equation}
{\cal D}_\text{weak}(t) = \frac{1}{\Delta^2} \, \frac{1}{R_2} \,
\frac{\langle \op{J}^2 \rangle}{\langle \op{X}^2 \rangle} =
\text{const.} \, , \label{weakD}
\end{equation}
i.e., the diffusion coefficient scales as $1/\Delta^2$, if the
other quantities $\langle \op{J}^2 \rangle$ and $\langle \op{X}^2
\rangle$ do not depend on $\Delta$. However, the underlying
assumption of constant rates may not be fulfilled, or at least
not for the projection onto a single observable. A possible
criterion for the validity of such a lowest order prediction
might be the independence from the number of observables, see
Sec.~\ref{Heisenberg}.
\\
In the limit of strong interactions a truncation to lowest order
appears to be not reasonable. On the one hand the interaction
strength is large as such, and on the other hand the relevant time
scales are short and the rates in all orders are explicitly
time-dependent here, e.g., in the non-Markovian limit. Moreover, in
that limit significant differences between the lowest order
predictions of the TCL and NZ approach are commonly expected due to
memory effects \cite{breuer2007}. But, as a consequence of the
explicit time-dependence of rates, a truncation to lowest order will
turn out to be justified anyhow.
\\
To this end consider the rates in Eq.~(\ref{r2}) and (\ref{r4}). For
sufficiently short times the \emph{integrands} can be treated as
approximately time-independent and the time-integrals can be easily
performed. Thus, in the case of short time scales, Eq.~(\ref{r2})
and (\ref{r4}) are approximated by
\begin{equation}
R_2(t) \approx r_2 \, t \, , \; r_2 = \frac{\langle \imath
[\op{J},\op{V}]^2 \rangle}{\langle J^2 \rangle} \, , \label{a2}
\end{equation}
\vspace{-0.5cm}
\begin{equation}
R_4(t) \approx r_4 \, \frac{t^3}{6} \, , \; r_4 = 3 \, r_2^2 -
\frac{\langle [ [\op{J},\op{V}], \op{V}]^2 \rangle}{\langle J^2
\rangle} \, , \label{a4}
\end{equation}
where the time-dependence of the $i$th order rate appears only as
the power $t^{i-1}$. Apparently, the approximations do not contain
any feature of the uncoupled system $\op{H}_0$, neither its
eigenvectors nor its eigenvalues. But $\op{H}_0$ still determines
the maximum time for the validity of these approximations. Or, in
other words, the approximations are valid, if the resulting decay
time $\tau$ becomes shorter than typical decay times in $\op{H}_0$,
see below.
\\
Now assume for the moment that a truncation to lowest order
\emph{was} indeed correct. Then Eq.~(13) directly predicts the
strictly Gaussian decay $C(t) = \exp(-\Delta^2 \, r_2 \, t^2/2) \,
\langle J^2 \rangle$ and for times above the decay time $\tau =
\sqrt{2}/(\Delta \, \sqrt{r_2})$ the diffusion coefficient in
Eq.~(\ref{D}) thus becomes
\begin{equation}
{\cal D}_\text{strong}(t) = \frac{1}{\Delta} \, \sqrt{\frac{\pi}{2
\, r_2}} \, \frac{\langle \op{J}^2 \rangle}{\langle \op{X}^2
\rangle} = \text{const.} \, , \label{strongD}
\end{equation}
i.e., the diffusion constant scales as $1/\Delta$ in that case, in
contrast to the former case of weak interactions. Such a prediction
is only correct, if higher order rates represent minor corrections
at the relevant time scale, i.e., up to the decay time $\tau$.
Hence, the ratio of the fourth to the second order rate may be
compared at this particular point in time. According to
Eqs.~(\ref{a2}) and (\ref{a4}), the ratio reads
\begin{equation}
\frac{\Delta^4 \, R_4(\tau)}{\Delta^2 \, R_2(\tau)} = \Delta^2 \,
\tau^2 \, \frac{r_4}{ 6 \, r_2} = 1 - \frac{\langle \op{J}^2
\rangle \, \langle [ [\op{J},\op{V}], \op{V}]^2 \rangle}{3 \,
\langle \imath [\op{J},\op{V}]^2 \rangle^2} \label{ratio}
\end{equation}
and does not depend on the interaction strength $\Delta$. The latter
independence from $\Delta$ also holds true for ratios with higher
order rates. Thus, the validity of the second order truncation for
strong interactions is connected to static expectation values,
involving exclusively the \emph{form} of the interaction and not its
strength as such. In a sense the situation is rather similar to weak
interactions.
\\
In principle the ratio in Eq.~(\ref{ratio}) can certainly be a large
negative number. But in many situations and also for all concrete
transport models in the following Sections the ratio turns out to be
rather close to zero. Due to such a small ratio, the relaxation of
the current is found to be well described in terms of the second
order prediction, at least up to the decay time. This restriction is
necessary, simply since the introduced approach does not allow to
describe zero-crossings of the current, see Eq.~(\ref{decay}). Such
a zero-crossing essentially requires the divergence of the whole
series expansion at this point in time. However, it will be
demonstrated that the relaxation of the current is described almost
exactly up to the (first) zero-crossing for strong interactions, if
fourth order corrections are taken into account. Particularly, the
case of strong interactions will be shown to include physically
relevant and not only pathologic situations.

% ----------------------------------------------------------------------------
%
% Section: Modular Quantum System
%
% ----------------------------------------------------------------------------

\section{Modular Quantum System}
\label{modular}

This Section will investigate the transport of a \emph{single}
excitation amongst local modules in an one-dimensional
configuration, as a first example for a concrete quantum system
\cite{gemmer2006,steinigeweg2007,steinigeweg2009-1}. This modular
quantum system is one of the few models which have been reliably
shown to exhibit purely diffusive transport. In fact, pure diffusion
has been found from different approaches and also the same
quantitative value of the diffusion coefficient has been deduced.
But all these approaches have focused on the case of weakly coupled
modules. The present investigation will address the so far
unexplored case of strong couplings.

To start with, each of the $N$ local modules features an identical
spectrum, consisting of $n$ equidistant levels in a band with the
width $\delta \epsilon$. Therefore the local Hamiltonian at the
position $r$ is given by $\op{h}_r = \sum_\mu \mu \, \delta
\epsilon/n \, |r,\mu\rangle\langle r,\mu|$ in the one-particle basis
$|r,\mu\rangle$. The next-neighbor coupling between two local
modules at the positions $r$ and $r+1$ is supposed to be
\begin{equation}
\Delta \, \op{v}_r = \Delta \sum_{\mu,\nu} c_{\mu,\nu} \,
|r,\mu\rangle\langle r+1,\nu| + \text{H.c.} \label{randomV}
\end{equation}
with the overall coupling strength $\Delta$. The $r$-independent
coefficients $c_{\mu,\nu}$ are complex, random, and uncorrelated
numbers: their real and imaginary part are both chosen corresponding
to a Gaussian distribution with the mean $0$ and the variance $1/2$.
Here, a \emph{single} realization of these coefficients is
considered (and not an ensemble average over different
realizations). The total Hamiltonian reads $\op{H} = \op{H}_0 +
\Delta \, \op{V}$, where $\op{H}_0$ denotes the sum of all local
Hamiltonians $\op{h}_r$ and $\op{V}$ represents respectively the sum
of all next-neighbor couplings $\op{v}_r$.
\\
Of particular interest is the probability for finding the excitation
somewhere within the $r$th local module. Such probabilities
correspond to local density operators of the form $\op{x}_r =
\sum_\mu | r,\mu \rangle\langle r,\mu|$. Their total sum is the
identity $1$ and a globally conserved quantity. Therefore, according
to the scheme in Sec.~\ref{current}, the definition of the
associated local current is given by $\op{j}_r = \imath
[\op{p}_r,\Delta \, \op{v}_r]$, yielding
\begin{equation}
\op{j}_r = \Delta \sum_{\mu,\nu} \imath \, c_{\mu,\nu} \,
|r,\mu\rangle\langle r+1,\nu| + \text{H.c.}
\end{equation}
with a form similar to $\op{v}_r$ in Eq.~(\ref{randomV}). The total
current $\op{J}$ does not commutate with $\op{H}$, $\op{H}_0$, or
$\op{V}$.

Although the modular quantum system is not meant to represent a
concrete physical situation, it may be viewed as a very simplified
model for a chain of coupled atoms or molecules. In that case the
hopping of the excitation from one module to another corresponds to
transport of energy. Alternatively, the modular quantum system may
be illustrated as an idealized model for non-interacting particles
on a more-dimensional lattice. In that case the hopping of the
excitation between modules corresponds to transport of particles
between chains or layers. There also is a relationship to the
three-dimensional Anderson model
\cite{anderson1958,steinigeweg2010-1}, even though the modular quantum
system does not contain any disorder, despite the random choice of
the coefficients $c_{\mu,\nu}$ in Eq.~(\ref{randomV}).

However, since the current $\op{J}$ does not commutate with the
uncoupled system $\op{H}_0$, the autocorrelation function $C(t)$ can
not be analyzed by the introduced approach in Sec.~\ref{lowest}, if
the coupling strength $\Delta$ is small. But for such small
$\Delta$ the dynamical behavior of the autocorrelation function
$C(t)$ can be analyzed in a different way. To this end consider the
total Hamiltonian $\op{H} = \op{H_0} + \Delta \, \op{V}$ for the
case of sufficiently small $\Delta$. In that case the eigensystem
of $\op{H}$ is essentially given by $\op{H}_0$, i.e., the
eigenvectors are $|r,\mu \rangle$ and the eigenvalues are $\mu \,
\delta\epsilon/n$. Obviously, $C(t)$ can be determined more or less
exactly: Since the spectrum features the width $\delta\epsilon$, the
autocorrelation function $C(t)$ fully decays at a first time scale
$\tau_0 \sim \pi/\delta\epsilon$ and eventually recurs completely at a
second time scale $T_0 = 2\pi \, n /\delta \epsilon$, simply due to
the equidistant levels of the spectrum. But, within the possibly
wide time window between these time scales, $C(t)$ remains zero and
thus the diffusion constant ${\cal D}(t)$ according to Eq.~(\ref{D})
becomes constant. Concretely, the diffusion constant reads
\cite{gemmer2006}
\begin{equation}
{\cal D}_\text{weak}(t) = \frac{\pi}{\delta \epsilon} \, \frac{\langle
\op{J}^2 \rangle}{\langle \op{X}^2 \rangle} = \frac{2 \pi \, \Delta^2
\, n}{\delta \epsilon} = \text{const.}
\end{equation}
for sufficiently many levels in the spectrum. The scaling factor
$\Delta^2$ results from $\langle J^2 \rangle = 2 \Delta^2 n$. This
result coincides with Fermi's Golden Rule and has been obtained
already from different approaches
\cite{gemmer2006,steinigeweg2007,steinigeweg2009-1}.

% ------------------------------- FIGURE2 ------------------------------------
\begin{figure}[tb]
\centering
\includegraphics[width=0.8\columnwidth]{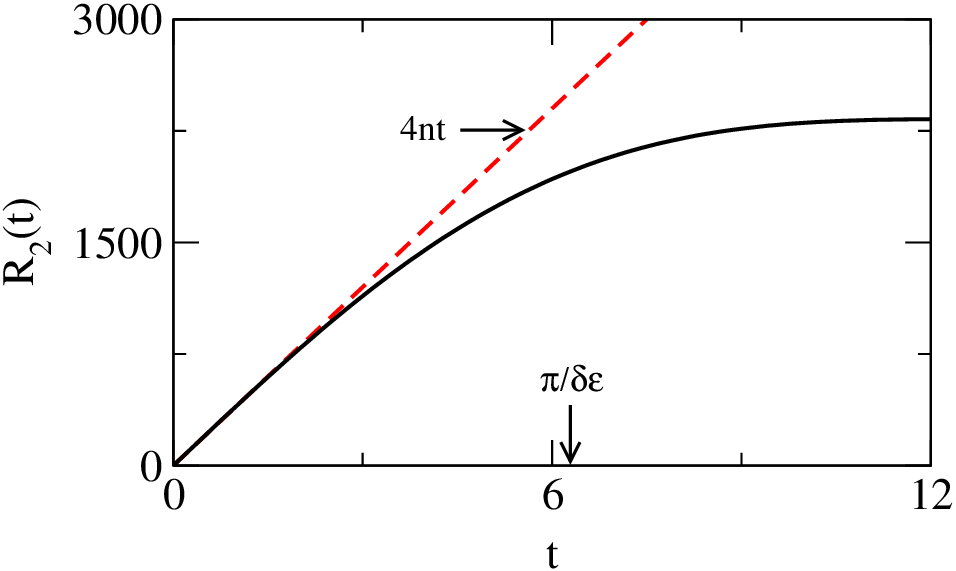}
\caption{(color online) The second order decay rate $R_2(t)$ for the
excitation current in the modular quantum system (black, solid
curve). Parameters: $n = 100$ and $\delta \epsilon = 0.5$. This
decay rate is well described by the linear approximation $4 \, n \,
t$ at a time scale below $\pi/\delta\epsilon$ (red, dashed line).}
\label{R_modular}
\end{figure}
%-----------------------------------------------------------------------------

\begin{figure}[tb]
\centering
\includegraphics[width=0.8\columnwidth]{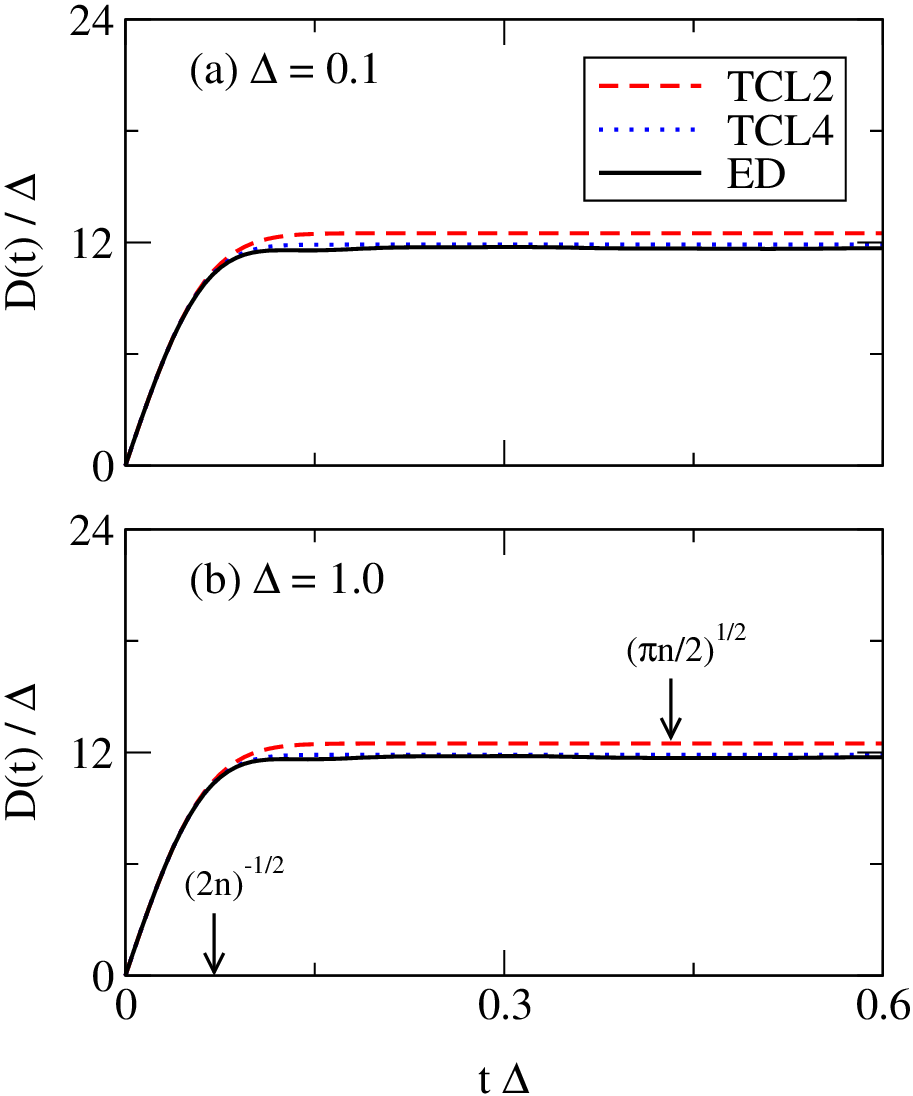}
\caption{(color online) The time-dependent diffusion constant ${\cal
D}(t)$ for the modular quantum system with strong interactions (a)
$\Delta = 0.1$, (b) $\Delta = 1.0$. Parameters: $N=1000$, $n = 100$,
and $\delta \epsilon=0.5$. The theoretical predictions according to
TCL2 (dashed, red curves) are already in good agreement with the
numerical results from ED (solid, black curves), while minor
corrections are given by the approximation of TCL4 (dotted, blue
curves). As predicted by theory, ${\cal D}(t)$ takes on a constant
value ${\cal D} \propto \Delta$ at intermediate times above the
relaxation time $\tau \propto 1/\Delta$.} \label{D_modular}
\end{figure}

In the so far unexplored case of strong couplings the approach in
Sec.~\ref{lowest} becomes applicable. The application only requires
that the decay due to $\Delta \, \op{V}$ proceeds much faster than
the decay due to $\op{H}_0$, i.e., $C(t)$ has to decay at a time
scale $\tau < \tau_0$. If $\Delta$ is large, this requirement is
naturally fulfilled. In fact, at such a time scale below $\tau_0$ no
correlation function w.r.t.~$\op{H}_0$ has decayed yet. Hence, the
second order rate in Eq.~(\ref{r2}) can be well described by the
linear approximation in Eq.~(\ref{a2}), see Fig.~\ref{R_modular}.
The use of $\langle \imath[\op{J},\op{V}]^2 \rangle = 8 \, \Delta^2
\, n^2$ and $\langle J^2 \rangle = 2 \, \Delta^2 \, n$ concretely
leads to $r_2 = 4 \, n$. The resulting second order prediction for
the diffusion constant in Eq.~(\ref{strongD}) consequently is
\begin{equation}
{\cal D}_\text{strong}(t) = \frac{1}{\Delta} \, \sqrt{\frac{\pi}{8 \,
n}} \, \frac{\langle \op{J}^2 \rangle}{\langle \op{X}^2 \rangle} =
\sqrt{\frac{\pi \, n}{2}} \, \Delta = \text{const.}
\end{equation}
above the relaxation time $\tau = 1/(\sqrt{2 \,n} \, \Delta)$. In
Fig.~\ref{D_modular} the analytical second order prediction is
compared with the direct numerical result for the diffusion
coefficient by the use of ED. Since for the modular quantum system
the linear growth of the Hilbert space can be compensated by the
translation symmetry, rather many modules are treatable. Apparently,
the agreement between the second order prediction and numerics is
very good, despite the limit of strong interactions. The minor
deviations can be further reduced by fourth order corrections in
terms of the cubic approximation in Eq.~(\ref{a4}). Concretely, the
use of $\langle [[\op{J}, \op{V}],\op{V}]^2 \rangle = 80 \, \Delta^2
\, n^3$ leads to $r_4 = 8 \, n^2$. With such fourth order
corrections the agreement in Fig.~\ref{D_modular} becomes excellent.
Remarkably, the corrections are small, because the ratio in
Eq.~(\ref{ratio}) takes on the value $1/6$.

% ------------------------------- FIGURE3 ------------------------------------
%----------------------------------------------------------------------------

% ----------------------------------------------------------------------------
%
% Section: Heisenberg Chain
%
% ----------------------------------------------------------------------------

\section{Heisenberg Chain}
\label{Heisenberg}

In this Section spin transport in the Heisenberg spin chain will be
investigated, as another concrete example for an interacting
\emph{many}-particle quantum system, going beyond the simplified
single-particle model in the last Section. This investigation will
firstly focus on a certain generalization of the standard Heisenberg
spin-$1/2$ chain, taking into account the effect of anisotropic
nearest and next-to-nearest neighbor interactions. The Hamiltonian
concretely reads $\op{H} = \op{H}_0 + \Delta \op{V}$, where the
operators $\op{H}_0$ and $\op{V}$ are given by
\begin{equation}
\op{H}_0 = \Jec \sum_r \op{s}_r^x \op{s}_{r+1}^x + \op{s}_r^y
\op{s}_{r+1}^y \, , \label{H0_Heisenberg}
\end{equation}
\vspace{-0.5cm}
\begin{equation}
\op{V} = \Jec \sum_r \op{s}_r^z \op{s}_{r+1}^z +
\frac{\Delta_2}{\Delta} \, \op{s}_r^z \op{s}_{r+2}^z \, .
\label{V_Heisenberg}
\end{equation}
Here, $\Jec$ is the exchange coupling constant, $\Delta$ refers to
the anisotropy parameter, the matrix $\op{s}_r^i$ represents the
$i$th component of the spin-$1/2$ operator at site $r$, and the
parameter $\Delta_2$ specifies the strength of an additional
next-to-nearest neighbor $zz$-interaction. For $\Delta_2 = 0$ the
Hamiltonian obviously reduces to the usual anisotropic Heisenberg
spin-$1/2$ chain (XXZ model).
\\
The transport of spin or magnetization corresponds to local density
operators $\op{x}_r = \op{s}_r^z$. Their sum is $\op{S}^z$ and a
globally conserved quantity. Therefore, according to the scheme in
Sec.~\ref{current}, the associated current can be written in the
well-known form (see, e.g., the reviews
\onlinecite{zotos2003,heidrichmeisner2007})
\begin{equation}
\op{J} = \Jec \sum_r \op{s}_r^x \op{s}_{r+1}^y - \op{s}_r^y
\op{s}_{r+1}^x \label{J_Heisenberg}
\end{equation}
and commutates with $\op{H}_0$. In particular, $\langle \op{J}^2
\rangle = \Jec^2 N/8$ and $\langle X^2 \rangle = N/4$, where the
trace operation is performed over the full $2^N$-dimensional Hilbert
space. In fact, the following investigation will not be restricted
to a specific $M$-subspace of $\op{S}^z$. However, the dominant
contribution to transport stems from the largest subspaces around
$M=0$ (`half filling'). The dynamics in these subspaces \emph{can} be
diffusive, while the dynamics in the subspaces with $|M| \gg 0$
(`dilute filling') is expected to be ballistic, see below.

Since the eigensystem of $\op{H}_0$ is indispensable for the
application of the introduced approach in Sec.~\ref{lowest}, it is
convenient to perform the Jordan-Wigner transformation onto
spin-less fermions\cite{jordan1928} firstly and an additional Fourier
transformation afterwards. The operators $\op{H}_0$, $\op{V}$, and
$\op{J}$ in Eqs.~(\ref{H0_Heisenberg})--(\ref{J_Heisenberg}) can then
be rewritten as\cite{mahan2000}
\begin{eqnarray}
&& \! \op{H}_0 = \Jec \sum_k \epsilon_k \, \op{n}_k \, , \, \epsilon_k
= \cos(k) \, , \nonumber \\
&& \! \op{V} = \frac{\Jec}{N} \sum_{k,l,q} v_q \, \op{a}_{k+q}^\dagger
a_{l-q}^\dagger \op{a}_l \op{a}_k \, , \, v_q = e^{\imath q} +
\frac{\Delta_2}{\Delta} e^{\imath 2 q} \, , \nonumber \\
&& \! \op{J} = \Jec \sum_k j_k \, \op{n}_k \, , \, j_k = \frac{\partial
\epsilon_k}{\partial k} = -\sin(k) \, . \label{JW}
\end{eqnarray}
Here, $\op{n}_k = \op{a}_k^\dagger \op{a}_k$ denotes the particle
number operator for a spin-less fermion with the momentum $k = 2\pi
i/N$ and is written as the product of respective creation and
annihilation operators. In this picture, $\op{H}_0$ describes the
dispersion $\epsilon_k$ of non-interacting particles, while $\op{V}$
is the interaction between two particles, located at nearest or
next-to-nearest sites, and $\op{J}$ plays the role of a particle
current. Since $\op{H}_0$ and $\op{J}$ are both diagonal, $J$ is
strictly preserved in the absence of $\op{V}$ and also in the
one-particle subspace. As a consequence a single particle propagates
ballistically.
\\
If additional next-to-nearest neighbor $xx$-/$yy$-terms are added to
Eq.~(\ref{H0_Heisenberg}) [and hence to Eq.~(\ref{J_Heisenberg})],
such a picture can not be established: $\op{H}_0$ does not become
diagonal by the use of the Jordan-Wigner transformation and $\op{J}$
does not commute with $\op{H}_0$. But these facts do not imply that
the approach as such becomes not applicable. $\op{H}_0$ can be
diagonalized at least numerically and, for large $\Delta$, the
commutation of $\op{J}$ with $\op{H}_0$ is not required in the strict
sense. Obviously, the situation is similar to the modular quantum
system in Sec.~\ref{modular}. However, it turns out that the decay of
$\op{J}$ w.r.t.~$\op{H}_0$ is comparatively fast. This fast decay
restricts the applicability of the approach to \emph{very} large
$\Delta$, i.e., close to the less interesting Ising limit. Thus, a
situation with next-to-nearest neighbor $xx$-/$yy$-terms will not be
discussed further.
\\
For operators of the form in Eq.~(\ref{JW}) an exact analytical
formula for the second order decay rate $R_2(t)$ in Eq.~(\ref{r2})
can be derived. In fact, the derivation of such a formula only
requires the concrete evaluation of the expectation value $\langle
[\op{n}_k,\op{V}]_\text{I}(t) \, [\op{n}_l,\op{V}] \rangle$ in the
interaction picture, i.e., w.r.t.~$\op{H}_0$. Even though the
concrete evaluation appears to be a straightforward task at the
first view, it turns out to be a both subtle and lengthy
calculation. Nevertheless, after such a calculation the latter
expectation value can be finally given as
\begin{eqnarray}
&& \frac{\langle \imath [\op{n}_k,\op{V}]_\text{I}(t) \, \imath
[\op{n}_l,\op{V}] \rangle}{\langle \op{J}^2 \rangle} = \frac{2}{N^3}
\sum_q \nonumber \\[-0.15cm]
&& \!\! \delta_{k,l} \sum_m (\tilde{v}_{k-q} - \tilde{v}_{m-q})^2
\cos[(\epsilon_k + \epsilon_m - \epsilon_{k+m-q} - \epsilon_q) \, t]
\nonumber \\[-0.15cm]
&& \!\! + (\tilde{v}_{k-q} - \tilde{v}_{l-q})^2 \cos[(\epsilon_k +
\epsilon_l - \epsilon_{k+l-q} - \epsilon_q) \, t] \nonumber \\[0.1cm]
&& \!\! -2 \, (\tilde{v}_{k-q} - \tilde{v}_{l-k})^2 \cos[(\epsilon_q
+ \epsilon_l - \epsilon_{q+l-k} - \epsilon_k) \, t]
\end{eqnarray}
with $\tilde{v}_k = \text{Re} \, v_k$. Consequently, for a linear
combination of the form $\op{A}_i = \sum_k a_k^i \, \op{n}_k$ one
finds
\begin{eqnarray}
R_2^{i,j}(t) \!\! &=& \!\! \int_0^t \! \text{d}t_1 \, \frac{\langle
\imath [\op{A}_i,\op{V}]_\text{I}(t_1) \, \imath [\op{A}_j,\op{V}]
\rangle}{\langle \op{J}^2 \rangle} \nonumber \\[0.1cm]
&=& \!\! \frac{2}{N^3} \sum_{k,l,q} (a_k^i a_k^j + a_k^i a_l^j - 2 \,
a_q^i a_l^j) \, (\tilde{v}_{k-q} - \tilde{v}_{l-q})^2 \nonumber \\
&\cdot& \!\! \int_0^t \! \text{d}t_1 \, \cos[(\epsilon_k +
\epsilon_l - \epsilon_{k+l-q} - \epsilon_q) \, t_1] \, ,
\label{r2_Heisenberg}
\end{eqnarray}
including the second order decay rate $R_2(t)$ for $i=j=1$ and
$a_k^1 = j_k$. The more general notation in
Eq.~(\ref{r2_Heisenberg}) will become useful later. Because this
equation only involves a sum over three momenta $k$,$l$, and $q$, it
can be evaluated numerically for several thousands of spins and,
say, in the thermodynamic limit. Concretely, $N = 2000$ will be
chosen in the following. For that choice finite size effects do not
appear at time scales up to $500/\Jec$. For instance, such finite size
effects occur at time scales on the order of $10/\Jec$, if $N = 20$ is
chosen \cite{steinigeweg2010-2}, e.g., the maximum number of spins
for ED.

\subsection{The case $\Delta_2 = \Delta$}

% ------------------------------- FIGURE4 ------------------------------------
\begin{figure}[tb]
\centering
\includegraphics[width=0.8\columnwidth]{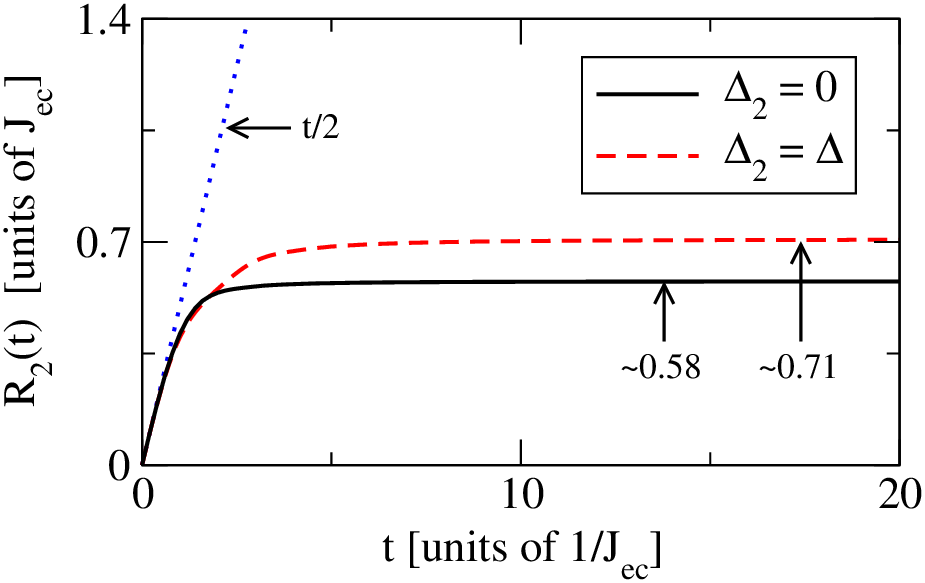}
\caption{(color online) The second order decay rate $R_2(t)$ for the
spin current in the anisotropic Heisenberg chain without additional
terms (black, solid curve) and with an additional next-to-nearest
neighbor $zz$-interaction of the same strength $\Delta_2 = \Delta$
(red, dashed curve). Parameters: $N = 2000$ [possible due to
Eq.~(\ref{r2_Heisenberg})]. These decay rates are well described by
the linear approximation $t/2$ at a time scale below $1/\Jec$ (blue,
dotted line). Finite size effects do not occur at time scales up to
$500/\Jec$.} \label{R_Heisenberg}
\end{figure}
%-----------------------------------------------------------------------------

% ------------------------------- FIGURE5 ------------------------------------
\begin{figure}[tb]
\centering
\includegraphics[width=0.8\columnwidth]{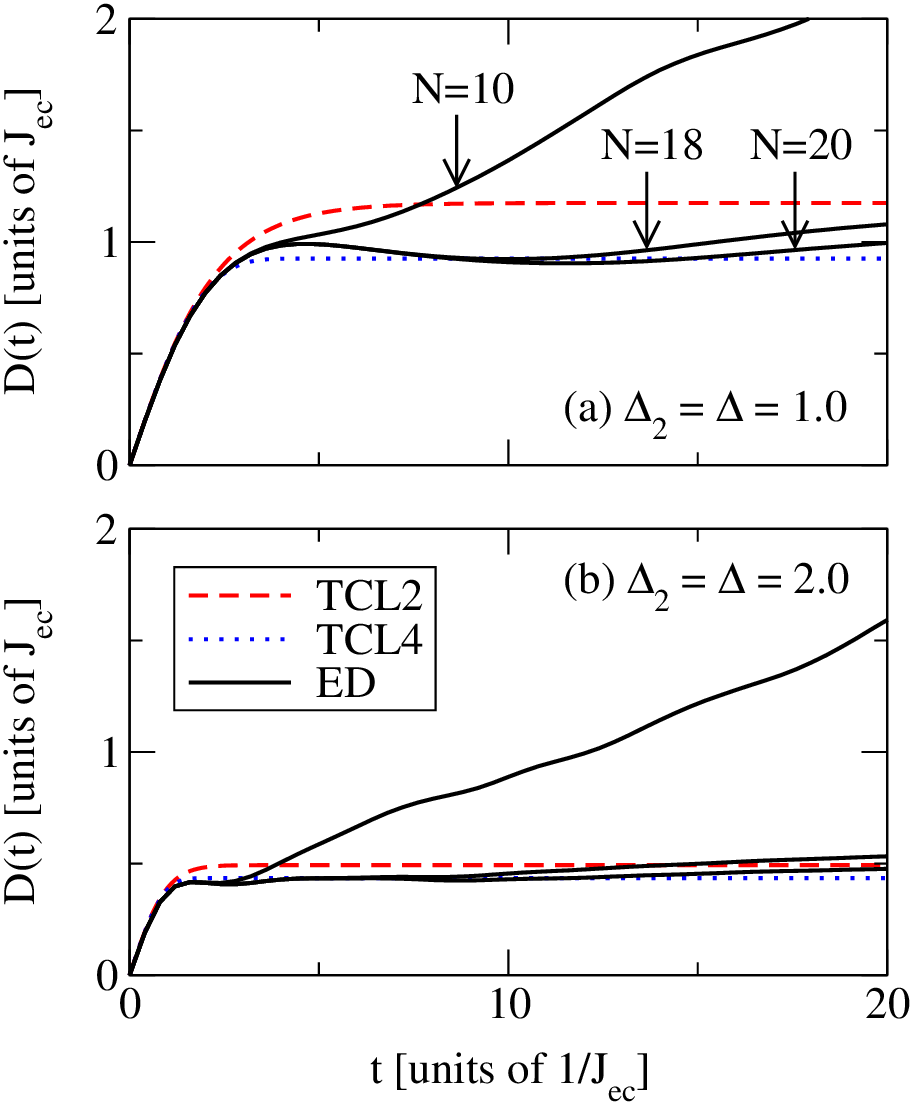}
\caption{(color online) The time-dependent diffusion constant ${\cal
D}(t)$ for spin transport in the anisotropic Heisenberg chain with
an additional next-to-nearest neighbor $zz$-interaction of the same
strength $\Delta_2 = \Delta$ for large anisotropy parameter (a)
$\Delta = 1.0$ and (b) $\Delta = 2.0$. The theoretical predictions
of TCL2 for $N= 2000$ (red, dashed curves) are already in good
agreement with the numerical results from ED for $N = 10$, $18$, and
$20$ (black, solid curves). Additional corrections on the order of
$20\%$ are given by the approximation of TCL4 (blue, dotted curves).
In (a) the approximation of TCL4 is a slight overestimation.}
\label{D_Heisenberg}
\end{figure}
%-----------------------------------------------------------------------------

% ------------------------------- FIGURE6 ------------------------------------
\begin{figure}[tb]
\centering
\includegraphics[width=0.8\columnwidth]{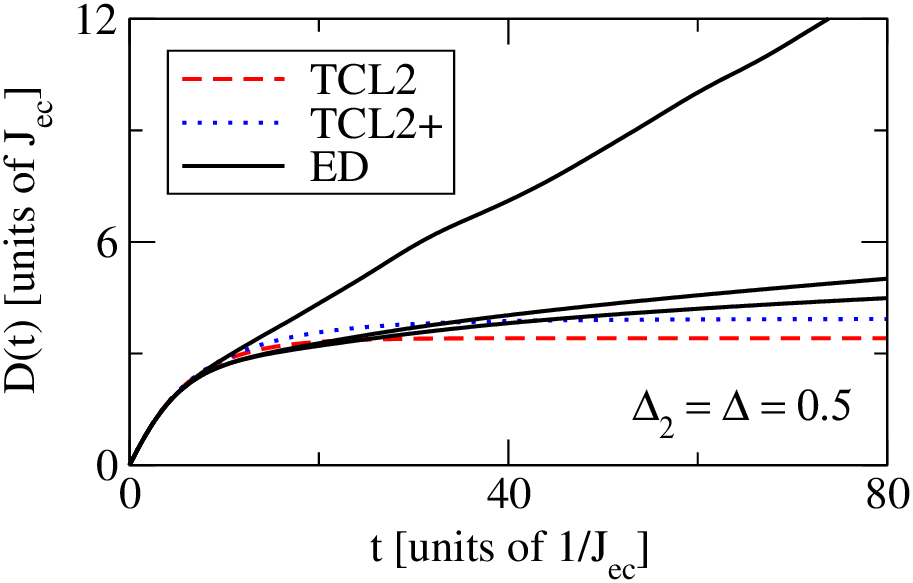}
\caption{(color online) The time-dependent diffusion constant ${\cal
D}(t)$ for spin transport in the anisotropic Heisenberg chain with
an additional next-to-nearest neighbor $zz$-interaction of the same
strength $\Delta_2 = \Delta = 0.5$. The theoretical prediction of
TCL2 for $N = 2000$ (red, dashed curve) is consistent with the
numerical results from ED for $N = 10$, $18$, and $20$ (black, solid
curves). Another projection with more observables leads to a
modified prediction TCL2+ for $N=200$ close to TCL2 (blue, dotted
curve).} \label{D_Heisenberg_2}
\end{figure}
%-----------------------------------------------------------------------------

First, the case of additional next-to-nearest neighbor
$zz$-interactions of the same strength may be discussed in detail,
i.e., $\Delta_2 = \Delta$. This particular case appears to be less
controversial, since Drude weights are commonly expected to vanish
in the thermodynamic limit due to non-integrability
\cite{zotos1996,narozhny1998,rabson2004}, at least if
$\Delta_2$ ($= \Delta$) does not become too small. For the
case $\Delta_2 = \Delta$ the second order decay rate $R_2(t)$ in
Fig.~\ref{R_Heisenberg} indeed takes on a form, as already
considered in Sec.~\ref{lowest}: $R_2(t)$ firstly increases linearly
at short time scales below $1/\Jec$ and then becomes constant at longer
time scales. Thus, the second order predictions for the diffusion
coefficient $\cal D$ can directly be formulated according to
Eqs.~(\ref{weakD}) and (\ref{strongD}). By the use of $R_2 \approx
0.71 \, \Jec$ from Fig.~\ref{R_Heisenberg} and $r_2 = 1/2 \, \Jec^2$
these predictions read
\begin{equation}
{\cal D}_\text{weak} \approx \frac{\Jec}{\Delta^2} \, 0.70 \, , \,
{\cal D}_\text{strong} = \frac{\Jec}{\Delta} \, \frac{\sqrt{\pi}}{2}
\, . \label{D_Delta2}
\end{equation}
For anisotropy parameters above $\Delta \sim 1$ the decay of the
current autocorrelation function takes place at a short time scale,
i.e., where $R_2(t)$ scales more or less linearly with time. Since
Drude weights for such $\Delta$ are already sufficiently small for
$N \sim 20$, a direct comparison with the numerical results from ED
becomes possible here, as shown in Fig.~\ref{D_Heisenberg}. By the
use of the exact $R_2(t)$ in Fig.~\ref{R_Heisenberg} the second
order predictions for ${\cal D}(t)$ are already in good agreement
with ED. Additional corrections on the order of $20\%$ are given by
the fourth order approximation in Eq.~(\ref{a4}). For $\Delta = 1$
the latter approximation for short times seems to be a slight
overestimation and may not be used further for smaller $\Delta$,
i.e., when the relevant time scales for the decay of the current
autocorrelation function become longer.
\\
Since Drude weights always become dominant for small $\Delta$ and
\emph{finite} systems, a direct comparison between the second order
prediction and the numerical results from ED is difficult in that
case. But for $\Delta = 0.5$ theory and numerics are at least
consistent, see Fig.~\ref{D_Heisenberg_2}. Because the fourth order
approximation is not available, the validity of the second order
prediction may be confirmed by its independence from the chosen
projection. To this end the projection may be extended to the full
\emph{diagonal} space, consisting of linear combinations $\op{A}_i$
of particle number operators $\op{n}_k$. As outlined in
Sec.~\ref{lowest}, such a extension of the projection yields a rate
matrix equation, namely,
\begin{equation}
\frac{\text{d}}{\text{d}t} \, \langle \op{J}(t) \, \op{A}_i \rangle
= - \Delta^2 \sum_j R_2^{i,j}(t) \, \langle \op{J}(t) \, A_j
\rangle
\end{equation}
with decay rates $R_2^{i,j}(t)$ according to
Eq.~(\ref{r2_Heisenberg}). This rate matrix equation can be solved
numerically by standard algorithms for, e.g., $N = 200$. This
modified prediction of, say, TCL2+ for $\Delta = 0.5$ turns out to
be rather close to the original prediction of TCL2, see
Fig.~\ref{D_Heisenberg_2}. Because both predictions become identical
for larger $\Delta$, TCL2+ is not indicated explicitly in
Fig.~\ref{D_Heisenberg}. However, TCL2 and TCL2+ begin to differ
significantly for smaller $\Delta$ and the validity of
Eq.~(\ref{D_Delta2}) in the limit of \emph{very} weak interactions
is questionable without the consideration of higher order decay
rates.

\subsection{The case $\Delta_2 = 0$}

% ------------------------------- FIGURE7 ------------------------------------
\begin{figure}[tb]
\centering
\includegraphics[width=0.8\columnwidth]{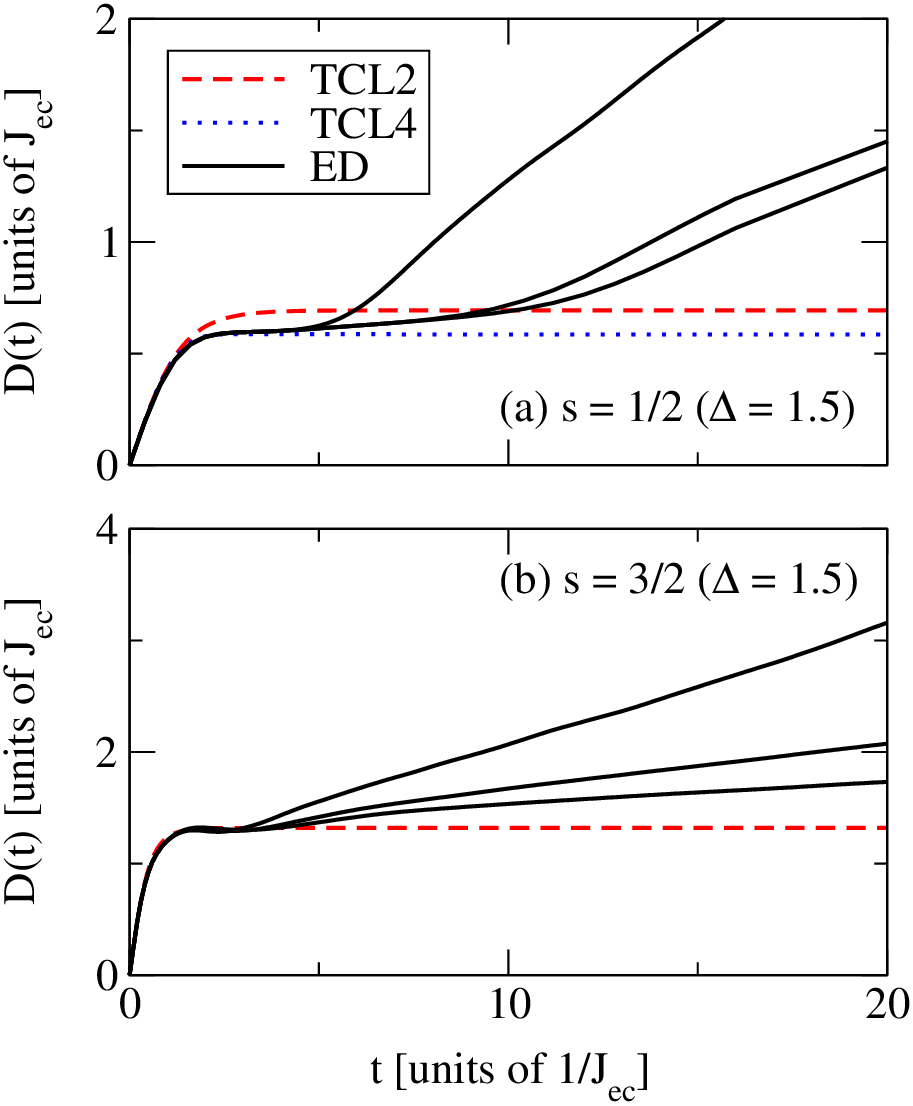}
\caption{(color online) The time-dependent diffusion constant ${\cal
D}(t)$ for spin transport in the anisotropic Heisenberg chain for
the anisotropy $\Delta = 1.5$ and the spin quantum numbers (a)
$s=1/2$ and (b) $s = 3/2$. The numerical results from ED (black,
solid curves) are shown for (a) $N = 10, 18, 20$ and (b) $N = 7, 8,
9$. The TCL2 predictions (red, dashed curves) are based on the data
in Fig.~\ref{R_Heisenberg} and Ref.~\onlinecite{steinigeweg2010-2},
respectively. No TCL4 corrections (blue, dotted curve) are required
in (b).} \label{D_Heisenberg_3}
\end{figure}
%-----------------------------------------------------------------------------

The case $\Delta_2 = 0$ is rather controversial due to the
integrability of the Hamiltonian in terms of the Bethe Ansatz
\cite{bethe1931}. In particular, for the isotropic point $\Delta = 1$,
there still is an unsettled debate about the finiteness of Drude weights
in the thermodynamic limit: While Drude weights are widely expected to
vanish for $\Delta > 1$ \cite{heidrichmeisner2003,prelovsek2004}, 
they may already become zero for $\Delta = 1$
\cite{zotos1999,benz2005,grossjohann2010}
but definitely non-zero for all $0 < \Delta < 1$ \cite{prosen2011}, see
also Ref.~\onlinecite{sirker2009}. However, in the present approach the
situation is found to be similar to the previous case $\Delta_2 = \Delta$,
see Fig.~\ref{R_Heisenberg}. The second order decay rate $R_2(t)$ is
almost the same, i.e., with a slightly reduced value $R_2 \approx 0.58
\, \Jec$. Remarkably, the latter value can already be \emph{supposed} on the
basis of $N \sim 20$, i.e., $\op{H}_0$ in Eq.~(\ref{JW}) contains only
$10$ different energies \cite{steinigeweg2010-2}. Due to
Fig.~\ref{R_Heisenberg}, the second order prediction for
${\cal D}_\text{strong}$ in Eq.~(\ref{D_Delta2}) remains unchanged
and ${\cal D}_\text{weak}$ becomes
\begin{equation}
{\cal D}_\text{weak} \approx \frac{\Jec}{\Delta^2} \, 0.86 \, .
\label{D_Delta}
\end{equation}
But this prediction has to be considered carefully, since it depends
on the chosen projection in the limit of very weak interactions,
analogously to the case $\Delta_2 = \Delta$. In fact,
significant differences between TCL2 and TCL2+ occur already for
$\Delta \sim 1$. Nevertheless, for larger $\Delta$ the TCL2
prediction is again found to be in good agreement with the numerical
results from ED, see Fig.~\ref{D_Heisenberg_3} (a). The additional
incorporation of the TCL4 approximation in Eq.~(\ref{a4}) goes beyond
Ref.~\onlinecite{steinigeweg2010-2} and explains the reported deviations
from ED and other approaches \cite{prelovsek2004,michel2008,prosen2009},
see also the perturbative approach in Ref.~\onlinecite{znidaric2011}.

\subsection{The case $s > 1/2$ ($\Delta_2 = 0$)}

For spin quantum numbers $s > 1/2$ the operators $\op{H}_0$,
$\op{V}$, and $\op{J}$ in
Eqs.~(\ref{H0_Heisenberg})--(\ref{J_Heisenberg}) are formally
identical. But in that case $\op{H}_0$ can not be brought into
diagonal form by the use of the Jordan-Wigner transformation. The
latter lack of a diagonal form is not a substantial drawback, since
$\op{J}$ does not commutate with $\op{H}_0$ for $s > 1/2$. Thus, the
investigation is anyway restricted to the limit of strong
interactions, similarly to the modular quantum system in
Sec.~\ref{modular}. Such an investigation has already been done in
detail in Ref.~\onlinecite{steinigeweg2010-2}. For illustration,
however, an example for $s = 3/2$ and $\Delta = 1.5$ is shown in
Fig.~\ref{D_Heisenberg_3} (b). In a sense it is intriguing to see
that the agreement between the theoretical prediction of TCL2 and
the numerical result from ED is best, if $[ \op{J}, \op{H}_0 ] \neq
0$, see also Fig.~\ref{D_modular}. It is worth to mention that the
investigation in Ref.~\onlinecite{steinigeweg2010-2} suggests that,
at high temperatures, the diffusion constant scales with the spin
quantum number as ${\cal D}_\text{strong} \propto \sqrt{s(s+1)}$,
see also
Refs.~\onlinecite{huber1969-1,huber1969-2,karadamoglou2004}. This
scaling supports classical simulations at high temperatures
\cite{mueller1988,gerling1989}.

\subsection{Alternating magnetic field ($\Delta_2 = 0$)}

% ------------------------------- FIGURE8 ------------------------------------
\begin{figure}[tb]
\centering
\includegraphics[width=0.8\columnwidth]{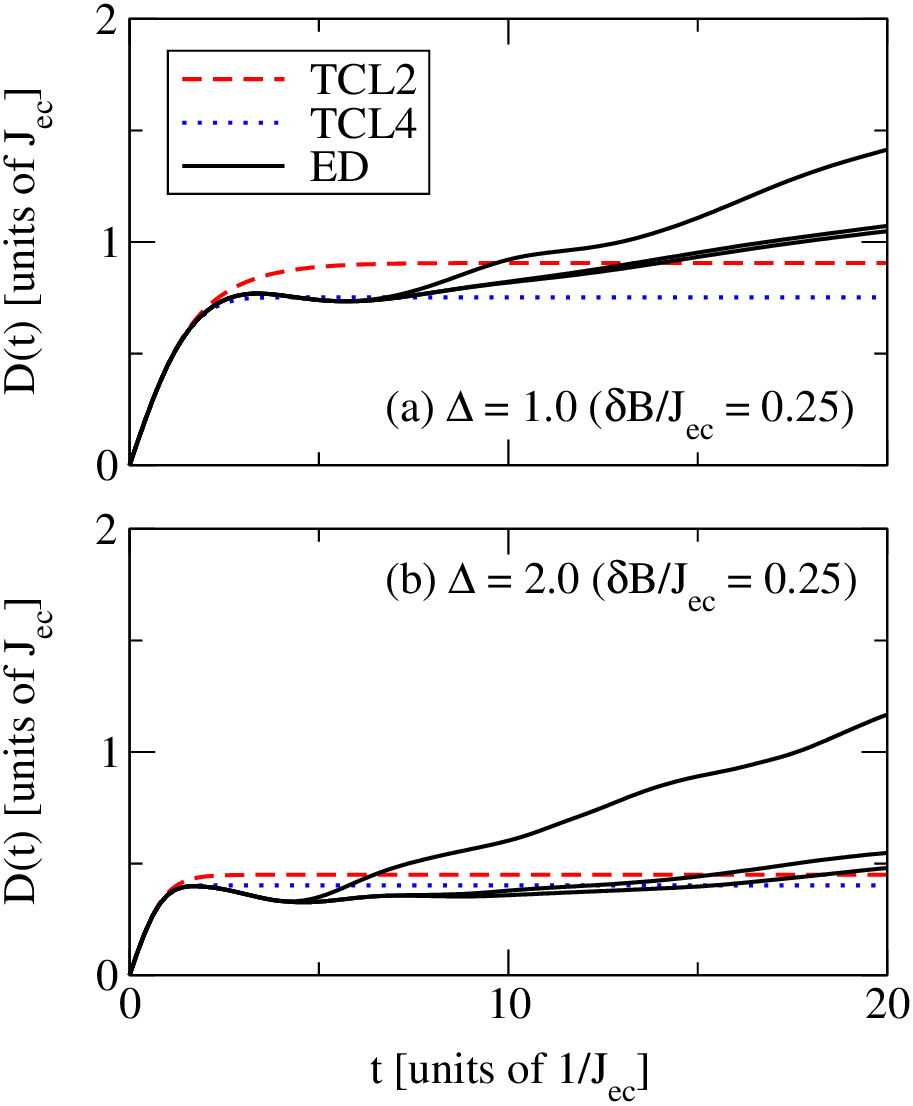}
\caption{(color online) The time-dependent diffusion constant ${\cal
D}(t)$ for spin transport in the anisotropic Heisenberg chain in the
presence of an alternating magnetic field $\delta B/\Jec = 0.25$ for
(a) $\Delta = 1.0$ and (b) $\Delta = 2.0$. The numerical results
from ED (black, solid curves) are shown for $N = 10$, $16$, and
$18$. The theoretical predictions of TCL2 (red, dashed curves) are
also based on numerical data for $N = 18$, even though TCL2 does not
depend on $N$ any further.} \label{D_Heisenberg_4}
\end{figure}
%-----------------------------------------------------------------------------

So far, the investigation at high temperatures does not depend on
the presence of a homogenous magnetic field $B$ in $z$-direction,
i.e., it does not change for an additional Zeeman term $\op{H}_B = B
\op{S}^z$ in Eqs.~(\ref{H0_Heisenberg}) and (\ref{V_Heisenberg}).
But the situation changes in the presence of an inhomogeneous
magnetic field
\cite{avishai2002,santos2004,santos2008,karahalios2009}, e.g., for
an alternating sequence
\cite{michel2008,prosen2009,steinigeweg2009-2,prosen2010}
\begin{equation}
\op{H}_B = \sum_r \, \frac{B +(-1)^r \, \delta B}{\Delta} \, \op{s}_r^z
\, , \label{alternating}
\end{equation}
where $\delta B$ denotes the strength of the alternation. Since
$[\op{J}, \op{H}_B] \neq 0$, a magnetic field of this form
represents an additional scattering mechanism and has to be added to
$\op{V}$ in Eq.~(\ref{V_Heisenberg}). In principle the alternating
magnetic field in Eq.~(\ref{alternating}) may also be written in the
representation of spin-less fermions and an exact analytical formula
for the second order decay rate $R_2(t)$ may again be derived. But,
because Eq.~(\ref{alternating}) obviously is no two-particle
interaction, $R_2(t)$ is not of the form in
Eq.~(\ref{r2_Heisenberg}). However, the case of an alternating
magnetic field is primarily considered here in order to demonstrate
potential difficulties of the approach at hand. For that reason the
decay rate $R_2(t)$ is directly evaluated numerically for $N \sim
18$ by the use of ED \cite{jung2006,jung2007}, e.g., finite systems
of this size are sufficient for the limit of \emph{strong}
interactions \cite{steinigeweg2010-2}. In Fig.~\ref{D_Heisenberg_4}
the resulting TCL2 prediction for ${\cal D}(t)$ is shown for $\delta
B/\Jec$ = 0.25 and different $\Delta$. Apparently, there still is a
good agreement with the numerical results from ED apart from the,
say, oscillation in Fig.~\ref{D_Heisenberg_4} (b). This oscillation
takes place after a zero-crossing of the underlying current
autocorrelation function \cite{comment} and is therefore not captured
by the present approach. Because the amplitude of such oscillations
is known to increase with the strength of the alternation
\cite{steinigeweg2009-2}, the TCL2 approach yields only meaningful
predictions for not too large $\delta B$.

% ----------------------------------------------------------------------------
%
% Section: Ising Chain
%
% ----------------------------------------------------------------------------

\section{Ising Chain}
\label{Ising}

This Section will deal with another concrete quantum system which
also allows to clarify potential difficulties of the approach at
hand. This quantum system is an Ising spin-$1/2$ chain in the
presence of a, say, tilted magnetic field. The Hamiltonian
concretely reads $\op{H} = \op{H}_0 + B_z \, \op{V}$, where the
operators $\op{H}_0$ and \op{V} are given by
\cite{mejiamonasterio2005,mejiamonasterio2007,prosen2009,steinigeweg2009-2}
\begin{equation}
\op{H}_0 = B_x \, \op{S}^x + \Jec \sum_r \op{s}_r^z \op{s}_{r+1}^z \, ,
\, \op{V} = \op{S}^z \, ,
\end{equation}
where $B_i$ and $\op{S}^i$ are the $i$th component of the magnetic
field ${\bf B}$ and total spin $\op{{\bf S}}$, respectively. For
instance, one might think of a magnetic field which was originally
in line with the $z$-direction and has been rotated about the
$y$-axis with the angle $\alpha = \arctan(B_x/B_z)$.
\\
Since $[\op{S}^z, \op{H}] \neq 0$ for this model, spin or
magnetization is not a suitable transport quantity here. However,
energy is always an appropriate transport quantity and may be
investigated instead. In that case the corresponding local density
operators read
\begin{equation}
\op{x}_r = \frac{B_x}{2} (\op{s}_r^x + \op{s}_{r+1}^x) + \Jec \,
\op{s}_r^z \op{s}_{r+1}^z + \frac{B_z}{2} (\op{s}_r^z +
\op{s}_{r+1}^z) \, .
\end{equation}
Thus, according to the scheme in Sec.~\ref{current}, the associated
current is given by
\begin{equation}
\op{J} = \frac{\Jec \, B_x}{2} \sum_r (\op{s}_{r+2}^z - \op{s}_r^z) \,
\op{s}_{r+1}^y
\end{equation}
and $[\op{J}, \op{H}_0] = 0$. Particularly, $\langle J^2 \rangle = N
\Jec^2 B_x^2 / 32$ as well as $\langle X^2 \rangle = N (4 B_x^2 + \Jec^2
+ 4 B_z^2)/16$.

\begin{figure}[tb]
\centering
\includegraphics[width=0.8\columnwidth]{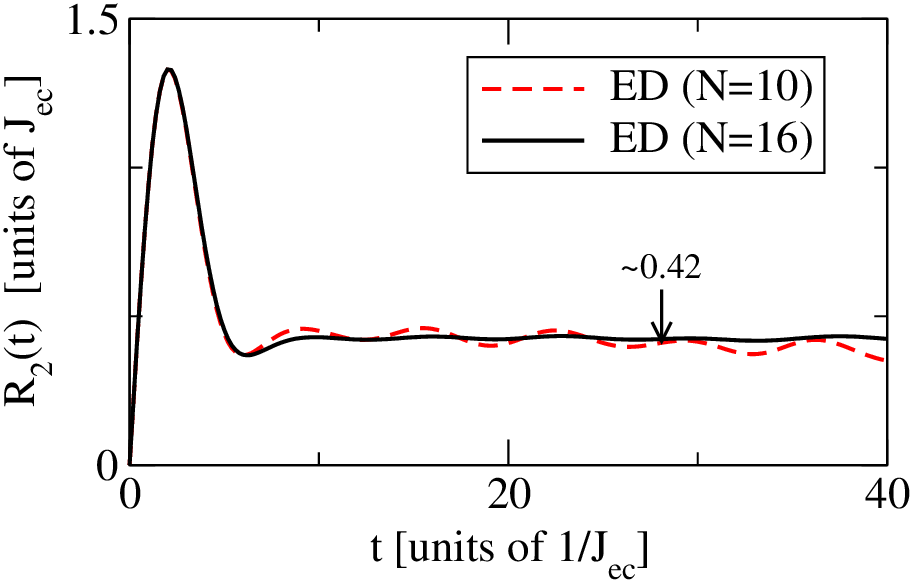}
\caption{(color online) The second order decay rate $R_2(t)$ for the
energy current in the Ising chain in the presence of a tilted
magnetic field for $N = 10$ and $16$. Parameters: $B_x/\Jec = 0.85$.}
\label{R_Ising}
\end{figure}

\begin{figure}[tb]
\centering
\includegraphics[width=0.8\columnwidth]{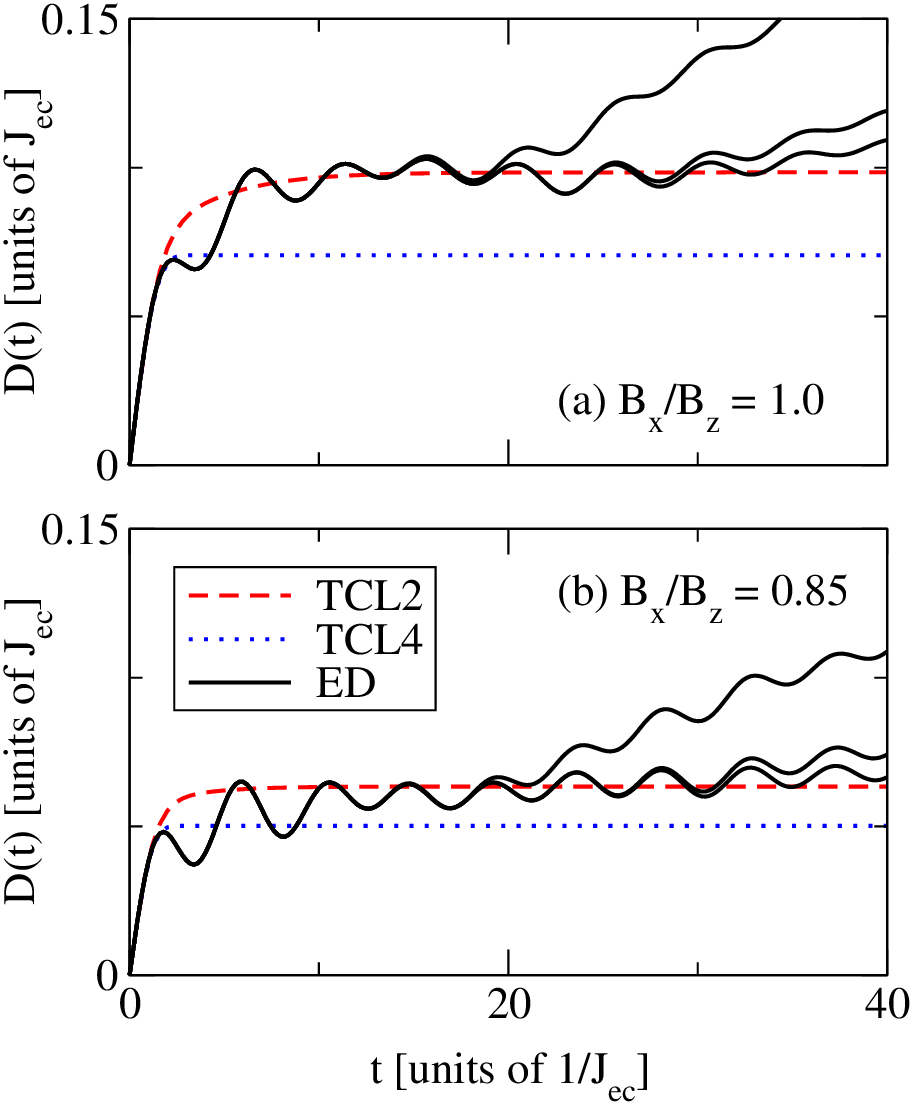}
\caption{(color online) The time-dependent diffusion constant ${\cal
D}(t)$ for energy transport in the Ising chain in the presence of a
tilted magnetic field for (a) $B_x/B_z = 1.0$ ($\alpha = 45^\circ$)
and (b) $B_x/B_z = 0.85$ ($\alpha \approx 40^\circ$). Parameters:
$B_x/\Jec = 0.85$. The numerical results from ED (black, solid curves)
are shown for $N = 10$, $14$, and $16$. The theoretical predictions
according to TCL2 (red, dashed curves) are also based on numerical
data for $N = 16$ in Fig.~\ref{R_Ising}.} \label{D_Ising}
\end{figure}

Because of the commutation of the operators $\op{J}$ and $\op{H}_0$
a second order prediction may be formulated for the limit of weak
`interactions', i.e., small $z$-components of the magnetic field. In
fact, direct numerics for $N \sim 16$ by the use of ED already
indicate a well-behaved second order decay rate $R_2(t)$, i.e.,
$R_2(t)$ appears to take on a constant value at long time scales,
see Fig.~\ref{R_Ising}. Although $N \sim 16$ is still far away from
the thermodynamic limit, the convergence with $N$ in
Fig.~\ref{R_Ising} seems to be at least rather indicative for a
constant value. However, since at short time scales $R_2(t)$ shows a
non-trivial dependence on time, such a dependence certainly occurs
also for higher order decay rates. In particular higher order decay
rates may not develop towards constant values at long time scales.
On that account a second order prediction has to be considered
carefully here, see Sec.~\ref{lowest}.
\\
Nevertheless, the convergence in Fig.~\ref{R_Ising} is sufficient
for \emph{strong} interactions \cite{steinigeweg2010-2}, analogously
to the previous case of an alternating magnetic field. The resulting
second order prediction for ${\cal D}(t)$ is shown in
Fig.~\ref{D_Ising} for equally large components of the magnetic
field, i.e., $B_x/B_z \sim 1$. The use of the fourth
order approximation again allows to correctly describe ${\cal D}(t)$
up to the (first) zero-crossing of the underlying current autocorrelation
function \cite{comment}. But, in contrast to the decrease in
Fig.~\ref{D_Heisenberg_4}, a renewed \emph{increase} of ${\cal
D}(t)$ emerges after this zero-crossing, resulting from partial
revivals of the current autocorrelation function, e.g., due to a
spectrum which gradually becomes closer to the equidistant levels of
the pure Ising model (with a magnetic field in $z$-direction). This
renewed increase of ${\cal D}(t)$ remarkably turns out to be
captured by the mere second order prediction. Therefore the present
example clearly illustrates that fourth order corrections improve
usually the description at short time scales but do not yield
necessarily to a better description at larger time scales, e.g.,
after a potential zero-crossing of the current autocorrelation
function.

% ------------------------------- FIGURE9 -------------------------------------
%------------------------------------------------------------------------------

% ------------------------------- FIGURE10 ------------------------------------
%-----------------------------------------------------------------------------

% ----------------------------------------------------------------------------
%
% Section: Summary and Conclusion
%
% ----------------------------------------------------------------------------

\section{Summary and Conclusion}
\label{summary}

The present paper has studied the decay of current autocorrelation
functions for quantum systems featuring strong `interactions'. In
this study the term interaction has referred to that part of the
Hamiltonian causing the (major) decay of the current. To this end an
appropriate perturbation theory in the interaction strength has been
introduced at first, namely, by an application of the TCL projection
operator technique. For the addressed case of strong interactions
the quality of a truncation to lowest order has been demonstrated to
depend on the form of the interaction and not on its strength as
such. By the use of the introduced perturbation theory the diffusion
coefficient has been evaluated afterwards for a variety of transport
quantities in concrete quantum systems. This evaluation has been
started for excitation transport in the modular quantum system and
has been continued for spin and energy transport in several spin
chains. For all examples the lowest order prediction for the
diffusion constant has well agreed with the numerical results from
ED, even in the case of strong interactions. Remarkably, higher
order corrections have played a minor role.

The investigation has focused on high temperatures and
one-dimensional quantum systems so far. Both have been chosen here in
order to allow for a comparison with the numerical results from ED,
being more or less free of finite size effects for this particular
choice. However, the introduced perturbation theory is not
restricted to one dimension. But this perturbation theory probably
is restricted to high temperatures, at least in the addressed case
of strong interactions. Low temperatures confine the perturbation
theory in the form at hand to the case of weak interactions: Only
for that case an approximation of the statistical operator on the
basis of the uncoupled system is expected to be reliable at all.
Nevertheless, in the context of weak interactions, higher order
corrections probably play a more important role, similarly to high
temperatures. In any case further estimations for higher order
contributions certainly are desirable, not only for the projection
onto currents but also for the alternative projection onto
densities. These projections have already been applied successfully
for one-particle models \cite{steinigeweg2007,steinigeweg2010-1}.

\acknowledgments

The author sincerely thanks R.~Schnalle, C.~Bartsch, and J.~Gemmer
for fruitful discussions. Furthermore, the author gratefully
acknowledges financial support by the \emph{Deutsche
Forschungsgemeinschaft} through FOR 912.

%\bibliography{stronginteractions}

\begin{thebibliography}{52}%
\makeatletter
\providecommand \@ifxundefined [1]{%
 \ifx #1\undefined \expandafter \@firstoftwo
 \else \expandafter \@secondoftwo
\fi
}%
\providecommand \@ifnum [1]{%
 \ifnum #1\expandafter \@firstoftwo
 \else \expandafter \@secondoftwo
\fi
}%
\providecommand \enquote [1]{``#1''}%
\providecommand \bibnamefont  [1]{#1}%
\providecommand \bibfnamefont [1]{#1}%
\providecommand \citenamefont [1]{#1}%
\providecommand\href[0]{\@sanitize\@href}%
\providecommand\@href[1]{\endgroup\@@startlink{#1}\endgroup\@@href}%
\providecommand\@@href[1]{#1\@@endlink}%
\providecommand \@sanitize [0]{\begingroup\catcode`\&12\catcode`\#12\relax}%
\@ifxundefined \pdfoutput {\@firstoftwo}{%
 \@ifnum{\z@=\pdfoutput}{\@firstoftwo}{\@secondoftwo}%
}{%
 \providecommand\@@startlink[1]{\leavevmode}%
 \providecommand\@@endlink[0]{}%
}{%
 \providecommand\@@startlink[1]{%
  \leavevmode
  \pdfstartlink
   attr{/Border[0 0 1 ]/H/I/C[0 1 1]}%
   user{/Subtype/Link/A<</Type/Action/S/URI/URI(#1)>>}%
  \relax
 }%
 \providecommand\@@endlink[0]{\pdfendlink}%
}%
\providecommand \url  [0]{\begingroup\@sanitize \@url }%
\providecommand \@url [1]{\endgroup\@href {#1}{\urlprefix}}%
\providecommand \urlprefix [0]{URL }%
\providecommand \Eprint[0]{\href }%
\@ifxundefined \urlstyle {%
  \providecommand \doi [1]{doi:\discretionary{}{}{}#1}%
}{%
  \providecommand \doi [0]{doi:\discretionary{}{}{}\begingroup
  \urlstyle{rm}\Url }%
}%
\providecommand \doibase [0]{http://dx.doi.org/}%
\providecommand \Doi[1]{\href{\doibase#1}}%
\providecommand \bibAnnote [3]{%
  \BibitemShut{#1}%
  \begin{quotation}\noindent
    \textsc{Key:}\ #2\\\textsc{Annotation:}\ #3%
  \end{quotation}%
}%
\providecommand \bibAnnoteFile [2]{%
  \IfFileExists{#2}{\bibAnnote {#1} {#2} {\input{#2}}}{}%
}%
\providecommand \typeout [0]{\immediate \write \m@ne }%
\providecommand \selectlanguage [0]{\@gobble}%
\providecommand \bibinfo [0]{\@secondoftwo}%
\providecommand \bibfield [0]{\@secondoftwo}%
\providecommand \translation [1]{[#1]}%
\providecommand \BibitemOpen[0]{}%
\providecommand \bibitemStop [0]{}%
\providecommand \bibitemNoStop [0]{.\EOS\space}%
\providecommand \EOS [0]{\spacefactor3000\relax}%
\providecommand \BibitemShut [1]{\csname bibitem#1\endcsname}%
%</preamble>
\bibitem{kubo1991}%
  \BibitemOpen
  \bibfield{author}{%
  \bibinfo {author} {\bibnamefont{{R. Kubo}}}, \bibinfo {author}
  {\bibnamefont{{M. Yokota}}},\ and\ \bibinfo {author} {\bibnamefont{{S.
  Hashtisume}}},\ }%
  \emph{\bibinfo {title} {Statistical {P}hysics {II}: {N}onequilibrium
  {S}tatistical {M}echanics}},\ \bibinfo {edition} {2nd}\ ed.,\ Solid State
  Sciences\ (\bibinfo {publisher} {Springer, New York},\ \bibinfo {year}
  {1991})%
  \bibAnnoteFile{NoStop}{kubo1991}%
\bibitem{mahan2000}%
  \BibitemOpen
  \bibfield{author}{%
  \bibinfo {author} {\bibfnamefont{G.~D.}\ \bibnamefont{Mahan}},\ }%
  \emph{\bibinfo {title} {Many {P}article {P}hysics}},\ \bibinfo {edition}
  {3rd}\ ed.,\ Physics of {S}olids and {L}iquids\ (\bibinfo {publisher}
  {Springer, New York},\ \bibinfo {year} {2000})%
  \bibAnnoteFile{NoStop}{mahan2000}%
\bibitem{sirker2009}%
  \BibitemOpen
  \bibfield{author}{%
  \bibinfo {author} {\bibnamefont{{J. Sirker}}}, \bibinfo {author}
  {\bibnamefont{{R. G. Pereira}}},\ and\ \bibinfo {author} {\bibnamefont{{I.
  Affleck}}},\ }%
  \bibfield{journal}{%
  \bibinfo {journal} {Phys. Rev. Lett.}\ }%
  \textbf{\bibinfo {volume} {103}},\ \bibinfo {pages} {216602} (\bibinfo {year}
  {2009})%
  \bibAnnoteFile{NoStop}{sirker2009}%
\bibitem{nakajima1958}%
  \BibitemOpen
  \bibfield{author}{%
  \bibinfo {author} {\bibfnamefont{S.}~\bibnamefont{Nakajima}},\ }%
  \bibfield{journal}{%
  \bibinfo {journal} {Progr. Theor. Phys.}\ }%
  \textbf{\bibinfo {volume} {20}},\ \bibinfo {pages} {948} (\bibinfo {year}
  {1958})%
  \bibAnnoteFile{NoStop}{nakajima1958}%
\bibitem{zwanzig1960}%
  \BibitemOpen
  \bibfield{author}{%
  \bibinfo {author} {\bibfnamefont{R.}~\bibnamefont{Zwanzig}},\ }%
  \bibfield{journal}{%
  \bibinfo {journal} {J. Chem. Phys.}\ }%
  \textbf{\bibinfo {volume} {33}},\ \bibinfo {pages} {1338} (\bibinfo {year}
  {1960})%
  \bibAnnoteFile{NoStop}{zwanzig1960}%
\bibitem{mori1965}%
  \BibitemOpen
  \bibfield{author}{%
  \bibinfo {author} {\bibfnamefont{H.}~\bibnamefont{Mori}},\ }%
  \bibfield{journal}{%
  \bibinfo {journal} {Progr. Theor. Phys.}\ }%
  \textbf{\bibinfo {volume} {33}},\ \bibinfo {pages} {423} (\bibinfo {year}
  {1965})%
  \bibAnnoteFile{NoStop}{mori1965}%
\bibitem{forstner1975}%
  \BibitemOpen
  \bibfield{author}{%
  \bibinfo {author} {\bibfnamefont{D.}~\bibnamefont{Forster}},\ }%
  \emph{\bibinfo {title} {Hydrodynamic {F}luctuations, {B}roken {S}ymmetry, and
  {C}orrelation {F}unctions}}\ (\bibinfo {publisher} {Benjamin,
  Massachusetts},\ \bibinfo {year} {1975})%
  \bibAnnoteFile{NoStop}{forstner1975}%
\bibitem{chaturvedi1979}%
  \BibitemOpen
  \bibfield{author}{%
  \bibinfo {author} {\bibnamefont{{S. Chaturvedi}}}\ and\ \bibinfo {author}
  {\bibnamefont{{F. Shibata}}},\ }%
  \bibfield{journal}{%
  \bibinfo {journal} {Z. Phys. B}\ }%
  \textbf{\bibinfo {volume} {35}},\ \bibinfo {pages} {297} (\bibinfo {year}
  {1979})%
  \bibAnnoteFile{NoStop}{chaturvedi1979}%
\bibitem{breuer2007}%
  \BibitemOpen
  \bibfield{author}{%
  \bibinfo {author} {\bibnamefont{{H.-P. Breuer}}}\ and\ \bibinfo {author}
  {\bibnamefont{{F. Petruccione}}},\ }%
  \emph{\bibinfo {title} {The {T}heory of {O}pen {Q}uantum {S}ystems}}\
  (\bibinfo {publisher} {Oxford University Press, New York},\ \bibinfo {year}
  {2007})%
  \bibAnnoteFile{NoStop}{breuer2007}%
\bibitem{mejiamonasterio2007}%
  \BibitemOpen
  \bibfield{author}{%
  \bibinfo {author} {\bibnamefont{{C. Mej\'{i}a-Monasterio}}}\ and\ \bibinfo
  {author} {\bibnamefont{{H. Wichterich}}},\ }%
  \bibfield{journal}{%
  \bibinfo {journal} {Eur. Phys. J. Spec. Top.}\ }%
  \textbf{\bibinfo {volume} {151}},\ \bibinfo {pages} {113} (\bibinfo {year}
  {2007})%
  \bibAnnoteFile{NoStop}{mejiamonasterio2007}%
\bibitem{michel2008}%
  \BibitemOpen
  \bibfield{author}{%
  \bibinfo {author} {\bibnamefont{{M. Michel}}}, \bibinfo {author}
  {\bibnamefont{{O. Hess}}}, \bibinfo {author} {\bibnamefont{{H.
  Wichterich}}},\ and\ \bibinfo {author} {\bibnamefont{{J. Gemmer}}},\ }%
  \bibfield{journal}{%
  \bibinfo {journal} {Phys. Rev. B}\ }%
  \textbf{\bibinfo {volume} {77}},\ \bibinfo {pages} {104303} (\bibinfo {year}
  {2008})%
  \bibAnnoteFile{NoStop}{michel2008}%
\bibitem{prosen2009}%
  \BibitemOpen
  \bibfield{author}{%
  \bibinfo {author} {\bibnamefont{{T. Prosen}}}\ and\ \bibinfo {author}
  {\bibnamefont{{M. \v{Z}nidari\v{c}}}},\ }%
  \bibfield{journal}{%
  \bibinfo {journal} {J. Stat. Mech.}\ }%
  \textbf{\bibinfo {volume} {2009}},\ \bibinfo {pages} {P02035} (\bibinfo
  {year} {2009})%
  \bibAnnoteFile{NoStop}{prosen2009}%
\bibitem{steinigeweg2009-1}%
  \BibitemOpen
  \bibfield{author}{%
  \bibinfo {author} {\bibnamefont{{R. Steinigeweg}}}, \bibinfo {author}
  {\bibnamefont{{M. Ogiewa}}},\ and\ \bibinfo {author} {\bibnamefont{{J.
  Gemmer}}},\ }%
  \bibfield{journal}{%
  \bibinfo {journal} {EPL}\ }%
  \textbf{\bibinfo {volume} {87}},\ \bibinfo {pages} {10002} (\bibinfo {year}
  {2009})%
  \bibAnnoteFile{NoStop}{steinigeweg2009-1}%
\bibitem{prosen2010}%
  \BibitemOpen
  \bibfield{author}{%
  \bibinfo {author} {\bibnamefont{{T. Prosen}}}\ and\ \bibinfo {author}
  {\bibnamefont{{M. \v{Z}nidari\v{c}}}},\ }%
  \bibfield{journal}{%
  \bibinfo {journal} {Phys. Rev. Lett.}\ }%
  \textbf{\bibinfo {volume} {105}},\ \bibinfo {pages} {060603} (\bibinfo {year}
  {2010})%
  \bibAnnoteFile{NoStop}{prosen2010}%
\bibitem{prosen2011}%
  \BibitemOpen
  \bibfield{author}{%
  \bibinfo {author} {\bibfnamefont{T.}~\bibnamefont{Prosen}},\ }%
  \bibfield{journal}{%
  \bibinfo {journal} {Phys. Rev. Lett.}\ }%
  \textbf{\bibinfo {volume} {106}},\ \bibinfo {pages} {217206} (\bibinfo {year}
  {2011})%
  \bibAnnoteFile{NoStop}{prosen2011}%
\bibitem{znidaric2011}%
  \BibitemOpen
  \bibfield{author}{%
  \bibinfo {author} {\bibfnamefont{M.}~\bibnamefont{\v{Z}nidari\v{c}}},\ }%
  \bibfield{journal}{%
  \bibinfo {journal} {Phys. Rev. Lett.}\ }%
  \textbf{\bibinfo {volume} {106}},\ \bibinfo {pages} {220601} (\bibinfo {year}
  {2011})%
  \bibAnnoteFile{NoStop}{znidaric2011}%
\bibitem{jung2006}%
  \BibitemOpen
  \bibfield{author}{%
  \bibinfo {author} {\bibnamefont{{P. Jung}}}, \bibinfo {author}
  {\bibnamefont{{R. W. Helmes}}},\ and\ \bibinfo {author} {\bibnamefont{{A.
  Rosch}}},\ }%
  \bibfield{journal}{%
  \bibinfo {journal} {Phys. Rev. Lett.}\ }%
  \textbf{\bibinfo {volume} {96}},\ \bibinfo {pages} {067202} (\bibinfo {year}
  {2006})%
  \bibAnnoteFile{NoStop}{jung2006}%
\bibitem{jung2007}%
  \BibitemOpen
  \bibfield{author}{%
  \bibinfo {author} {\bibnamefont{{P. Jung}}}\ and\ \bibinfo {author}
  {\bibnamefont{{A. Rosch}}},\ }%
  \bibfield{journal}{%
  \bibinfo {journal} {Phys. Rev. B}\ }%
  \textbf{\bibinfo {volume} {76}},\ \bibinfo {pages} {245108} (\bibinfo {year}
  {2007})%
  \bibAnnoteFile{NoStop}{jung2007}%
\bibitem{steinigeweg2010-2}%
  \BibitemOpen
  \bibfield{author}{%
  \bibinfo {author} {\bibnamefont{{R. Steinigeweg}}}\ and\ \bibinfo {author}
  {\bibnamefont{{R. Schnalle}}},\ }%
  \bibfield{journal}{%
  \bibinfo {journal} {Phys. Rev. E}\ }%
  \textbf{\bibinfo {volume} {82}},\ \bibinfo {pages} {040103(R)} (\bibinfo
  {year} {2010})%
  \bibAnnoteFile{NoStop}{steinigeweg2010-2}%
\bibitem{steinigeweg2007}%
  \BibitemOpen
  \bibfield{author}{%
  \bibinfo {author} {\bibnamefont{{R. Steinigeweg}}}, \bibinfo {author}
  {\bibnamefont{{H.-P. Breuer}}},\ and\ \bibinfo {author} {\bibnamefont{{J.
  Gemmer}}},\ }%
  \bibfield{journal}{%
  \bibinfo {journal} {Phys. Rev. Lett.}\ }%
  \textbf{\bibinfo {volume} {99}},\ \bibinfo {pages} {150601} (\bibinfo {year}
  {2007})%
  \bibAnnoteFile{NoStop}{steinigeweg2007}%
\bibitem{zotos2003}%
  \BibitemOpen
  \bibfield{author}{%
  \bibinfo {author} {\bibnamefont{{X. Zotos}}}\ and\ \bibinfo {author}
  {\bibnamefont{{P. Prelo\v{v}sek}}},\ }%
  \emph{\bibinfo {title} {Interacting {E}lectrons in {L}ow {D}imensions}},\
  Physics and {C}hemistry of {M}aterials with {L}ow-{D}imensional {S}tructures\
  (\bibinfo {publisher} {Kluwer Academic, Dordrecht},\ \bibinfo {year} {2004})%
  \bibAnnoteFile{NoStop}{zotos2003}%
\bibitem{heidrichmeisner2007}%
  \BibitemOpen
  \bibfield{author}{%
  \bibinfo {author} {\bibnamefont{{F. Heidrich-Meisner}}}, \bibinfo {author}
  {\bibnamefont{{A. Honecker}}},\ and\ \bibinfo {author} {\bibnamefont{{W.
  Brenig}}},\ }%
  \bibfield{journal}{%
  \bibinfo {journal} {Eur. Phys. J. Spec. Top.}\ }%
  \textbf{\bibinfo {volume} {151}},\ \bibinfo {pages} {135} (\bibinfo {year}
  {2007})%
  \bibAnnoteFile{NoStop}{heidrichmeisner2007}%
\bibitem{prelovsek2004}%
  \BibitemOpen
  \bibfield{author}{%
  \bibinfo {author} {\bibnamefont{{P. Prelov\v{s}ek}}}, \bibinfo {author}
  {\bibnamefont{{S. El Shawish}}}, \bibinfo {author} {\bibnamefont{{X.
  Zotos}}},\ and\ \bibinfo {author} {\bibnamefont{{M. Long}}},\ }%
  \bibfield{journal}{%
  \bibinfo {journal} {Phys. Rev. B}\ }%
  \textbf{\bibinfo {volume} {70}},\ \bibinfo {pages} {205129} (\bibinfo {year}
  {2004})%
  \bibAnnoteFile{NoStop}{prelovsek2004}%
\bibitem{huber1969-1}%
  \BibitemOpen
  \bibfield{author}{%
  \bibinfo {author} {\bibnamefont{{D. L. Huber}}}\ and\ \bibinfo {author}
  {\bibnamefont{{J. S. Semura}}},\ }%
  \bibfield{journal}{%
  \bibinfo {journal} {Phys. Rev.}\ }%
  \textbf{\bibinfo {volume} {182}},\ \bibinfo {pages} {602} (\bibinfo {year}
  {1969})%
  \bibAnnoteFile{NoStop}{huber1969-1}%
\bibitem{huber1969-2}%
  \BibitemOpen
  \bibfield{author}{%
  \bibinfo {author} {\bibnamefont{{D. L. Huber}}}, \bibinfo {author}
  {\bibnamefont{{J. S. Semura}}},\ and\ \bibinfo {author} {\bibnamefont{{C. G.
  Windsor}}},\ }%
  \bibfield{journal}{%
  \bibinfo {journal} {Phys. Rev.}\ }%
  \textbf{\bibinfo {volume} {186}},\ \bibinfo {pages} {534} (\bibinfo {year}
  {1969})%
  \bibAnnoteFile{NoStop}{huber1969-2}%
\bibitem{karadamoglou2004}%
  \BibitemOpen
  \bibfield{author}{%
  \bibinfo {author} {\bibnamefont{{J. Karadamoglou}}}\ and\ \bibinfo {author}
  {\bibnamefont{{X. Zotos}}},\ }%
  \bibfield{journal}{%
  \bibinfo {journal} {Phys. Rev. Lett.}\ }%
  \textbf{\bibinfo {volume} {93}},\ \bibinfo {pages} {177203} (\bibinfo {year}
  {2004})%
  \bibAnnoteFile{NoStop}{karadamoglou2004}%
\bibitem{steinigeweg2009-3}%
  \BibitemOpen
  \bibfield{author}{%
  \bibinfo {author} {\bibnamefont{{R. Steinigeweg}}}, \bibinfo {author}
  {\bibnamefont{{H. Wichterich}}},\ and\ \bibinfo {author} {\bibnamefont{{J.
  Gemmer}}},\ }%
  \bibfield{journal}{%
  \bibinfo {journal} {EPL}\ }%
  \textbf{\bibinfo {volume} {88}},\ \bibinfo {pages} {10004} (\bibinfo {year}
  {2009})%
  \bibAnnoteFile{NoStop}{steinigeweg2009-3}%
\bibitem{goldstein2006}%
  \BibitemOpen
  \bibfield{author}{%
  \bibinfo {author} {\bibnamefont{{S. Goldstein}}}, \bibinfo {author}
  {\bibnamefont{{J. L. Lebowitz}}}, \bibinfo {author} {\bibnamefont{{R.
  Tumulka}}},\ and\ \bibinfo {author} {\bibnamefont{{N. Zanghi}}},\ }%
  \bibfield{journal}{%
  \bibinfo {journal} {Phys. Rev. Lett.}\ }%
  \textbf{\bibinfo {volume} {96}},\ \bibinfo {pages} {050403} (\bibinfo {year}
  {2006})%
  \bibAnnoteFile{NoStop}{goldstein2006}%
\bibitem{popescu2006}%
  \BibitemOpen
  \bibfield{author}{%
  \bibinfo {author} {\bibnamefont{{S. Popescu}}}, \bibinfo {author}
  {\bibnamefont{{A. J. Short}}},\ and\ \bibinfo {author} {\bibnamefont{{A.
  Winter}}},\ }%
  \bibfield{journal}{%
  \bibinfo {journal} {Nature Phys.}\ }%
  \textbf{\bibinfo {volume} {2}},\ \bibinfo {pages} {754} (\bibinfo {year}
  {2006})%
  \bibAnnoteFile{NoStop}{popescu2006}%
\bibitem{reimann2007}%
  \BibitemOpen
  \bibfield{author}{%
  \bibinfo {author} {\bibnamefont{{P. Reimann}}},\ }%
  \bibfield{journal}{%
  \bibinfo {journal} {Phys. Rev. Lett.}\ }%
  \textbf{\bibinfo {volume} {99}},\ \bibinfo {pages} {160404} (\bibinfo {year}
  {2007})%
  \bibAnnoteFile{NoStop}{reimann2007}%
\bibitem{bartsch2009}%
  \BibitemOpen
  \bibfield{author}{%
  \bibinfo {author} {\bibnamefont{{C. Bartsch}}}\ and\ \bibinfo {author}
  {\bibnamefont{{J. Gemmer}}},\ }%
  \bibfield{journal}{%
  \bibinfo {journal} {Phys. Rev. Lett.}\ }%
  \textbf{\bibinfo {volume} {102}},\ \bibinfo {pages} {110403} (\bibinfo {year}
  {2009})%
  \bibAnnoteFile{NoStop}{bartsch2009}%
\bibitem{steinigeweg2009-2}%
  \BibitemOpen
  \bibfield{author}{%
  \bibinfo {author} {\bibnamefont{{R. Steinigeweg}}}\ and\ \bibinfo {author}
  {\bibnamefont{{J. Gemmer}}},\ }%
  \bibfield{journal}{%
  \bibinfo {journal} {Phys. Rev. B}\ }%
  \textbf{\bibinfo {volume} {80}},\ \bibinfo {pages} {184402} (\bibinfo {year}
  {2009})%
  \bibAnnoteFile{NoStop}{steinigeweg2009-2}%
\bibitem{gemmer2006}%
  \BibitemOpen
  \bibfield{author}{%
  \bibinfo {author} {\bibnamefont{{J. Gemmer}}}, \bibinfo {author}
  {\bibnamefont{{R. Steinigeweg}}},\ and\ \bibinfo {author} {\bibnamefont{{M.
  Michel}}},\ }%
  \bibfield{journal}{%
  \bibinfo {journal} {Phys. Rev. B}\ }%
  \textbf{\bibinfo {volume} {73}},\ \bibinfo {pages} {104302} (\bibinfo {year}
  {2006})%
  \bibAnnoteFile{NoStop}{gemmer2006}%
\bibitem{anderson1958}%
  \BibitemOpen
  \bibfield{author}{%
  \bibinfo {author} {\bibfnamefont{P.~W.}\ \bibnamefont{Anderson}},\ }%
  \bibfield{journal}{%
  \bibinfo {journal} {Phys. Rev.}\ }%
  \textbf{\bibinfo {volume} {109}},\ \bibinfo {pages} {1492} (\bibinfo {year}
  {1958})%
  \bibAnnoteFile{NoStop}{anderson1958}%
\bibitem{steinigeweg2010-1}%
  \BibitemOpen
  \bibfield{author}{%
  \bibinfo {author} {\bibnamefont{{R. Steinigeweg}}}, \bibinfo {author}
  {\bibnamefont{{H. Niemeyer}}},\ and\ \bibinfo {author} {\bibnamefont{{J.
  Gemmer}}},\ }%
  \bibfield{journal}{%
  \bibinfo {journal} {New J. Phys.}\ }%
  \textbf{\bibinfo {volume} {12}},\ \bibinfo {pages} {113001} (\bibinfo {year}
  {2010})%
  \bibAnnoteFile{NoStop}{steinigeweg2010-1}%
\bibitem{jordan1928}%
  \BibitemOpen
  \bibfield{author}{%
  \bibinfo {author} {\bibnamefont{{P. Jordan}}}\ and\ \bibinfo {author}
  {\bibnamefont{{E. Wigner}}},\ }%
  \bibfield{journal}{%
  \bibinfo {journal} {Z. Phys.}\ }%
  \textbf{\bibinfo {volume} {47}},\ \bibinfo {pages} {631} (\bibinfo {year}
  {1928})%
  \bibAnnoteFile{NoStop}{jordan1928}%
\bibitem{zotos1996}%
  \BibitemOpen
  \bibfield{author}{%
  \bibinfo {author} {\bibnamefont{{X. Zotos}}}\ and\ \bibinfo {author}
  {\bibnamefont{{P. Prelov\v{s}ek}}},\ }%
  \bibfield{journal}{%
  \bibinfo {journal} {Phys. Rev. B}\ }%
  \textbf{\bibinfo {volume} {53}},\ \bibinfo {pages} {983} (\bibinfo {year}
  {1996})%
  \bibAnnoteFile{NoStop}{zotos1996}%
\bibitem{narozhny1998}%
  \BibitemOpen
  \bibfield{author}{%
  \bibinfo {author} {\bibnamefont{{B. N. Narozhny}}}, \bibinfo {author}
  {\bibnamefont{{A. J. Millis}}},\ and\ \bibinfo {author} {\bibnamefont{{N.
  Andrei}}},\ }%
  \bibfield{journal}{%
  \bibinfo {journal} {Phys. Rev. B}\ }%
  \textbf{\bibinfo {volume} {58}},\ \bibinfo {pages} {2921(R)} (\bibinfo {year}
  {1998})%
  \bibAnnoteFile{NoStop}{narozhny1998}%
\bibitem{rabson2004}%
  \BibitemOpen
  \bibfield{author}{%
  \bibinfo {author} {\bibnamefont{{D. A. Rabson}}}, \bibinfo {author}
  {\bibnamefont{{B. N. Narozhny}}},\ and\ \bibinfo {author} {\bibnamefont{{A.
  J. Millis}}},\ }%
  \bibfield{journal}{%
  \bibinfo {journal} {Phys. Rev. B}\ }%
  \textbf{\bibinfo {volume} {69}},\ \bibinfo {pages} {054403} (\bibinfo {year}
  {2004})%
  \bibAnnoteFile{NoStop}{rabson2004}%
\bibitem{bethe1931}%
  \BibitemOpen
  \bibfield{author}{%
  \bibinfo {author} {\bibfnamefont{H.}~\bibnamefont{Bethe}},\ }%
  \bibfield{journal}{%
  \bibinfo {journal} {Z. Phys. A}\ }%
  \textbf{\bibinfo {volume} {71}},\ \bibinfo {pages} {205} (\bibinfo {year}
  {1931})%
  \bibAnnoteFile{NoStop}{bethe1931}%
\bibitem{heidrichmeisner2003}%
  \BibitemOpen
  \bibfield{author}{%
  \bibinfo {author} {\bibnamefont{{F. Heidrich-Meisner}}}, \bibinfo {author}
  {\bibnamefont{{A. Honecker}}}, \bibinfo {author} {\bibnamefont{{D. C.
  Cabra}}},\ and\ \bibinfo {author} {\bibnamefont{{W. Brenig}}},\ }%
  \bibfield{journal}{%
  \bibinfo {journal} {Phys. Rev. B}\ }%
  \textbf{\bibinfo {volume} {68}},\ \bibinfo {pages} {134436} (\bibinfo {year}
  {2003})%
  \bibAnnoteFile{NoStop}{heidrichmeisner2003}%
\bibitem{zotos1999}%
  \BibitemOpen
  \bibfield{author}{%
  \bibinfo {author} {\bibnamefont{{X. Zotos}}},\ }%
  \bibfield{journal}{%
  \bibinfo {journal} {Phys. Rev. Lett.}\ }%
  \textbf{\bibinfo {volume} {82}},\ \bibinfo {pages} {1764} (\bibinfo {year}
  {1999})%
  \bibAnnoteFile{NoStop}{zotos1999}%
\bibitem{benz2005}%
  \BibitemOpen
  \bibfield{author}{%
  \bibinfo {author} {\bibnamefont{{J. Benz}}}, \bibinfo {author}
  {\bibnamefont{{T. Fukui}}}, \bibinfo {author} {\bibnamefont{{A.
  Kl\"umper}}},\ and\ \bibinfo {author} {\bibnamefont{{C. Scheeren}}},\ }%
  \bibfield{journal}{%
  \bibinfo {journal} {J. Phys. Soc. Jpn.}\ }%
  \textbf{\bibinfo {volume} {74}},\ \bibinfo {pages} {181} (\bibinfo {year}
  {2005})%
  \bibAnnoteFile{NoStop}{benz2005}%
\bibitem{grossjohann2010}%
  \BibitemOpen
  \bibfield{author}{%
  \bibinfo {author} {\bibnamefont{{S. Grossjohann}}}\ and\ \bibinfo {author}
  {\bibnamefont{{W. Brenig}}},\ }%
  \bibfield{journal}{%
  \bibinfo {journal} {Phys. Rev. B}\ }%
  \textbf{\bibinfo {volume} {81}},\ \bibinfo {pages} {012404} (\bibinfo {year}
  {2010})%
  \bibAnnoteFile{NoStop}{grossjohann2010}%
\bibitem{mueller1988}%
  \BibitemOpen
  \bibfield{author}{%
  \bibinfo {author} {\bibfnamefont{G.}~\bibnamefont{M\"uller}},\ }%
  \bibfield{journal}{%
  \bibinfo {journal} {Phys. Rev. Lett.}\ }%
  \textbf{\bibinfo {volume} {60}},\ \bibinfo {pages} {2785} (\bibinfo {year}
  {1988})%
  \bibAnnoteFile{NoStop}{mueller1988}%
\bibitem{gerling1989}%
  \BibitemOpen
  \bibfield{author}{%
  \bibinfo {author} {\bibnamefont{{R. W. Gerling}}}\ and\ \bibinfo {author}
  {\bibnamefont{{D. P. Landau}}},\ }%
  \bibfield{journal}{%
  \bibinfo {journal} {Phys. Rev. Lett.}\ }%
  \textbf{\bibinfo {volume} {63}},\ \bibinfo {pages} {812} (\bibinfo {year}
  {1989})%
  \bibAnnoteFile{NoStop}{gerling1989}%
\bibitem{avishai2002}%
  \BibitemOpen
  \bibfield{author}{%
  \bibinfo {author} {\bibnamefont{{Y. Avishai}}}, \bibinfo {author}
  {\bibnamefont{{J. Richert}}},\ and\ \bibinfo {author} {\bibnamefont{{R.
  Berkovits}}},\ }%
  \bibfield{journal}{%
  \bibinfo {journal} {Phys. Rev. B}\ }%
  \textbf{\bibinfo {volume} {66}},\ \bibinfo {pages} {052416} (\bibinfo {year}
  {2002})%
  \bibAnnoteFile{NoStop}{avishai2002}%
\bibitem{santos2004}%
  \BibitemOpen
  \bibfield{author}{%
  \bibinfo {author} {\bibfnamefont{L.~F.}\ \bibnamefont{Santos}},\ }%
  \bibfield{journal}{%
  \bibinfo {journal} {J. Phys. A}\ }%
  \textbf{\bibinfo {volume} {37}},\ \bibinfo {pages} {4723} (\bibinfo {year}
  {2004})%
  \bibAnnoteFile{NoStop}{santos2004}%
\bibitem{santos2008}%
  \BibitemOpen
  \bibfield{author}{%
  \bibinfo {author} {\bibnamefont{{L. F. Santos}}},\ }%
  \bibfield{journal}{%
  \bibinfo {journal} {Phys. Rev. E}\ }%
  \textbf{\bibinfo {volume} {78}},\ \bibinfo {pages} {031125} (\bibinfo {year}
  {2008})%
  \bibAnnoteFile{NoStop}{santos2008}%
\bibitem{karahalios2009}%
  \BibitemOpen
  \bibfield{author}{%
  \bibinfo {author} {\bibnamefont{{A. Karahalios}}}, \bibinfo {author}
  {\bibnamefont{{A. Metavitsiadis}}}, \bibinfo {author} {\bibnamefont{{X.
  Zotos}}}, \bibinfo {author} {\bibnamefont{{A. Gorczyca}}},\ and\ \bibinfo
  {author} {\bibnamefont{{P. Prelov\v{s}ek}}},\ }%
  \bibfield{journal}{%
  \bibinfo {journal} {Phys. Rev. B}\ }%
  \textbf{\bibinfo {volume} {79}},\ \bibinfo {pages} {024425} (\bibinfo {year}
  {2009})%
  \bibAnnoteFile{NoStop}{karahalios2009}%
\bibitem{comment}%
  \BibitemOpen
  \bibinfo {note} {Since $\mathrm{d}/\mathrm{d}t \, {\cal D}(t) \propto C(t)$,
  a local extremum of ${\cal D}(t)$ implies a zero-crossing of $C(t)$.}%
  \bibAnnoteFile{Stop}{comment}%
\bibitem{mejiamonasterio2005}%
  \BibitemOpen
  \bibfield{author}{%
  \bibinfo {author} {\bibnamefont{{C. Mej\'{i}a-Monasterio}}}, \bibinfo
  {author} {\bibnamefont{{T. Prosen}}},\ and\ \bibinfo {author}
  {\bibnamefont{{G. Casati}}},\ }%
  \bibfield{journal}{%
  \bibinfo {journal} {Europhys. Lett.}\ }%
  \textbf{\bibinfo {volume} {72}},\ \bibinfo {pages} {520} (\bibinfo {year}
  {2005})%
  \bibAnnoteFile{NoStop}{mejiamonasterio2005}%
\end{thebibliography}

%

\end{document}